\documentclass[5p,times,twocolumn,authoryear]{elsarticle}

\usepackage{amsmath}
\usepackage{amssymb}
\usepackage{xcolor}
\usepackage{tikz}
\usetikzlibrary{arrows.meta,positioning,calc,patterns}
\usepackage{booktabs}
\usepackage{tabularx}
\usepackage{array}
\usepackage{multirow}
\usepackage{enumitem}
\usepackage{algorithm}
\usepackage{algpseudocode}
\usepackage[hidelinks]{hyperref}
\hypersetup{pdftitle={When "Benign" Is Unverified: Label-Signal Mismatch and Abstention-First Triage in Graph-Based Memory Forensics}, pdfcreator={}, pdfproducer={}}

\definecolor{benignblue}{HTML}{4C78A8}
\definecolor{malwarered}{HTML}{E45756}
\definecolor{gategrey}{HTML}{8C8C8C}

\newcommand{\codeus}{\char`\_\allowbreak}
\newcommand{\code}[1]{\textup{\texttt{\let\_\codeus #1}}}
\newcommand{\state}[1]{`#1'}
\newcommand{\eps}{\varepsilon}
\ifdefined\MARKEDCOPY
  \newcommand{\badd}[1]{\textcolor{blue}{#1}}
  \newcommand{\rfix}[1]{\textcolor{red}{#1}}
  \newcommand{\anote}[1]{\textcolor{red}{[#1]}}
\else
  \newcommand{\badd}[1]{#1}
  \newcommand{\rfix}[1]{#1}
  \newcommand{\anote}[1]{}
\fi
\algrenewcommand\algorithmicrequire{\textbf{Input:}}
\algrenewcommand\algorithmicensure{\textbf{Output:}}
\algrenewcommand\algorithmiccomment[1]{\hfill{\footnotesize$\triangleright$~#1}}

\newcommand{\doi}[1]{\href{https://doi.org/#1}{\nolinkurl{https://doi.org/#1}}}

\newtheorem{proposition}{Proposition}
\newtheorem{corollary}{Corollary}
\newdefinition{definition}{Definition}
\newproof{proof}{Proof}

\journal{Forensic Science International: Digital Investigation}

\begin{document}

\begin{frontmatter}

\title{When ``Benign'' Is \rfix{Unverified}: Label--Signal Mismatch and Abstention-First Triage in Graph-Based Memory Forensics}

\author[aff1]{Viraj Rajeshbhai Chudasama}
\ead{2435000@student.uwtsd.ac.uk}

\author[aff2]{Eduardo Palmieri}
\ead{e.a.palmieri@keele.ac.uk}

\author[aff1]{Dipo Dunsin}
\ead{d.dunsin@uwtsd.ac.uk}

\author[aff2]{Aisha Junejo}
\ead{a.junejo@keele.ac.uk}

\author[aff2,aff3]{Mohamed Chahine Ghanem\corref{cor1}}
\ead{m.ghanem@keele.ac.uk}
\cortext[cor1]{Corresponding author.}

\affiliation[aff1]{organization={Department of Applied Computing, IICL, University of Wales Trinity Saint David},
            city={{London}},
            postcode={E14 4HA},
            country={UK}}

\affiliation[aff2]{organization={School of Computer Science and Mathematics, Keele University},
            city={{Newcastle-under-Lyme}},
            postcode={ST5 5BG},
            country={UK}}

\affiliation[aff3]{organization={School of Computer Science and Informatics, University of Liverpool},
            city={{Liverpool}},
            postcode={L69 3BX},
            country={UK}}

\begin{abstract}
Machine-learning triage for memory forensics is usually validated on
corpora whose benign class is asserted by provenance and never verified. We
report a validation study of a graph-based triage pipeline for Windows memory
images: Volatility~3 output is compiled into typed heterogeneous graphs and
scored by a \rfix{temperature-scaled} GINE classifier and two one-class
encoders, fused with rule risk under an abstention gate.
\rfix{On MalVol-25 a transparent rule layer contradicts the benign label on
every control (15/15 and 18/18 flagged; at least 11 of 15 paired controls
rated HIGH or CRITICAL), and mean injection and command-and-control counts run
higher on controls than on infected captures. We call this rule--label
discordance label--signal mismatch; it shows the benign class is unverified,
not compromised.} Leave-one-family-out AUROC over matched pairs is provably
three-valued, so a reported mean and standard deviation identify the fold
composition. The learned model ($0.786\pm0.311$; 9/4/1\rfix{, scored
in-sample by a frozen checkpoint}) equals a logistic model over rule-derived
attributes ($0.786\pm0.410$; 11/0/3) in mean only, \rfix{and its significance
rests on dropping four saturated ties}. The fusion path sends $97.7\%$ of
snapshots to analyst review, $95.3\%$ with every gate disabled. Two public
feature-level corpora show the same defect: on CIC-MalMem-2022 every benign
row carries injected regions and one host-population feature reaches AUROC
\rfix{$0.997$} under leave-one-family-out validation; on the Sihwail et al.\
(2021) sample the injection indicator fires on $53\%$ of benign rows. We close
with seven reporting artefacts for learned triage on corpora whose ground
truth is asserted.
\end{abstract}


\end{frontmatter}

%% ==================================================================
\section{Introduction}
%% ==================================================================

Digital forensic tools are expected to have known, communicable error
rates~\citep{horsman2019tool,bolle2020evaluations}, and the field keeps
returning to validated methods and shared, documented corpora as the conditions
for progress~\citep{grajeda2017datasets,breitinger2023sharing}.
Machine-learning triage fits that expectation poorly. Memory-based malware
detectors are routinely reported at or near $99\%$ accuracy on
feature-engineered corpora whose benign class is balanced by
construction~\citep{carrier2022malmem,bozkir2021catch,dener2022malware}.
We reproduce that figure on CIC-MalMem-2022 in
Section~\ref{sec:external}, and show what it rests on. Such figures take
something for granted that is seldom audited, namely that the label attached to
a capture describes the state of the host. In malware corpora the ``benign''
class is often defined by provenance: a folder name, a snapshot taken before a
sample was detonated, a capture that nobody reported. A model can then be right
about the data and wrong about the world, and every property an examiner relies
on downstream (calibration, uncertainty, explanation) inherits the defect.

Windows memory forensics makes the problem easy to see. A memory image records
the volatile state of a live host: processes and their parentage, loaded
modules, virtual address descriptors (VADs), handles, threads, drivers and
network objects~\citep{ligh2014art,case2017memory,block2023imgmalfind,dehfouli2026memory}. The artefacts are
relational, which is the motivation for graph learning over
snapshots~\citep{song2018deepmem,bilot2024survey}. Even a host that has
never been infected leaves, in its memory image, a large volume of indicators
of the kind that malware also produces. Security products and other legitimate
software install API hooks in volumes that overwhelm manual
review~\citep{case2019hooktracer}. Heuristics of the \code{malfind} family flag
private executable memory that benign software also allocates, and at the same
time miss injected pages hidden from the VAD
view~\citep{ligh2014art,block2019pte}. Live acquisition is not atomic, and causal inconsistencies have been
shown to be the normal condition of a software-acquired memory dump, on Linux
and on Windows alike~\citep{pagani2019temporal,ottmann2024inconsistencies,rzepka2024causal,rzepka2025acquisition}.
Run an ordinary triage ruleset over a baseline image that is nominally
clean and it will often return a HIGH verdict, the second-highest of the four
risk bands (LOW, MEDIUM, HIGH, CRITICAL) that such rulesets use, and one that
would normally justify escalating the host to an examiner.

This paper takes that observation as its subject. We build a complete
pipeline, from Volatility~3 extraction to an abstention gate, and put it
through a validation protocol that asks \rfix{three questions. RQ1: does the
benign class of a paired memory corpus behave like a verified benign class
under a transparent, reproducible rule instrument? RQ2: does the learned
component add anything to the rules it is built on? RQ3: does the supervision
support the claims (calibration, coverage, generalisation) that would normally
be made from a learned triage model?} On this corpus the answer to \rfix{RQ1 is
no, and the answers to RQ2 and RQ3 are largely no}; the negative answers can be
measured and reproduced. We then ask whether the finding is peculiar to one
corpus, and repeat the audit on two public memory corpora that release
Volatility-derived features rather than images. The forensic question behind triage is the first one an
examiner asks of a host, ``what happened here, and is this machine
compromised''; our concern is whether a public forensic dataset can tell us how
well a tool answers it. \badd{Throughout, a rule-layer verdict is treated as
one imperfect witness against another, the provenance label; where the two
disagree we say that the benign label is unverified, and we do not infer from
the disagreement alone that the host was compromised.}

\paragraph{Contributions}
\begin{enumerate}[leftmargin=1.3em,itemsep=1pt,topsep=2pt]
\item We formalise \emph{label--signal mismatch} (mismatch rate, indicator gap
      and three physical causes) and turn it into a governance schema that
      records label difficulty per row, in machine-readable form
      (Section~\ref{sec:problem}). The schema makes the cleanliness of
      a benign class an auditable property of the corpus rather than an
      assumption of the paper, and it lets a strict training filter refuse
      the rows that fail the check.
\item We specify an \emph{abstention-first} architecture: a temperature-scaled
      GINE classifier and two one-class encoders fused with rule risk under a
      selective-classification gate, with attribute profiles that separate
      learned structure from rule-derived features (Section~\ref{sec:system}).
      The design answers a gap in learned forensic triage, where
      abstention is rarely reported as a result and learned scores are seldom
      separable from the rules that curated the training data: the rule layer
      stays usable on its own, every deferral is logged with the trigger that
      caused it, and the attribute profiles make it possible to show whether
      the learned component adds information or merely restates the rules.
      \rfix{Only the \code{full} profile was trained and scored in this study;
      the two leakage-controlled profiles are specified here and their
      evaluation is left as the first experiment for a verified corpus
      (Section~\ref{sec:discussion}).}
\item We give a construct-validity audit for matched-pair corpora, consisting
      of a leave-one-family-out (LOFO) ladder, a label-permutation null and
      $\eps$-tolerant pair diagnostics with exact intervals. We prove that
      two-graph folds make AUROC three-valued, so that a published mean and
      deviation reveal the fold composition, and we work out what a
      design of fourteen pairs can and cannot detect
      (Section~\ref{sec:construct}). The audit is needed because on a
      matched-pair corpus the customary summary, a mean AUROC over folds,
      cannot tell a model that orders pairs correctly from one that
      re-expresses the curation rules or from one that is simply tied, so a
      published figure can look significant while carrying no evidence of
      learning. \rfix{We also separate, for every rung of the ladder, an
      out-of-sample estimate (a model refitted without the held-out family)
      from an in-sample diagnostic (a frozen checkpoint that has seen every
      fold), and report which of the two each number is.}
\item We show by gate ablation that near-total abstention survives the removal
      of every uncertainty gate, quantify the degenerate score field behind it,
      and project the decisive path to realistic base rates
      (Sections~\ref{sec:binary}--\ref{sec:gate}).
\item We distil seven reporting artefacts for learned forensic triage on
      corpora with asserted ground truth (Section~\ref{sec:discussion}).
      The protocol turns caveats that are usually confined to a
      limitations section into artefacts that an author must produce and a
      reviewer can check, and it is what allows the same audit to be run on
      corpora that release features rather than images.
\item We run the audit on two public feature-level corpora,
      CIC-MalMem-2022 and the memory-artefact corpus of
      \citet{sihwail2021memory}, and find the benign-class problem in two
      further forms: a corpus on which a rule layer would flag every benign
      row while one host-population feature separates the classes under every
      hold-out, and one on which the injection indicator fires on half of the
      benign class (Section~\ref{sec:external}).
\end{enumerate}

We claim no detection advance: the pipeline does not detect malware
better than the rule layer it is built on, and the paper does not try to make
it do so, because a detection claim would require a benign class whose
cleanliness has been verified, which Section~\ref{sec:mismatch} shows this
corpus does not provide. \rfix{Selective} accuracy against manifest labels is
$0.537$ \badd{on the 41 decisive snapshots}, strict validation AUROC is
$0.438$, and the fusion path issues no decisive verdict at all. Reporting this
with the instrumentation that explains it is of more use to forensic practice
than a headline metric obtained by discarding the rows that make the problem
hard. \badd{The paper therefore distinguishes five things that a single
headline metric conflates: the validity of the dataset's labels, the validity
of the forensic instrument that contradicts them, the construct validity of
the evaluation design, the validity of the learned model, and the operational
usefulness of the triage it produces. Each is reported on its own evidence.}

%% ==================================================================
\section{Background and related work}
%% ==================================================================
This section places the study in four strands of prior work: what memory
forensics recovers and why its triage rules are noisy, the literature on
forensic datasets and ground truth, learned analysis of memory artefacts, and
work on evaluation validity, calibration and abstention. It closes by stating
the gaps that the paper addresses.

\paragraph{Memory forensics and its validation gap}
Volatile memory analysis recovers artefacts that never reach disk, such as
injected pages, unlinked process objects and short-lived
sockets~\citep{ligh2014art}. \citet{case2017memory} argue for scalable and
repeatable workflows. The literature also explains why a triage rule can be
evadable and noisy at once: injected code can be hidden from VAD-based
detection, the approach of \code{malfind} and similar tools, which walks
the virtual address descriptor tree of each process and flags private
regions that are executable,~\citep{block2019pte}, benign hooks are
everywhere~\citep{case2019hooktracer}, and acquisition is not
atomic~\citep{pagani2019temporal}. \citet{ottmann2024inconsistencies}
found some form of inconsistency in almost every second analysable dump of a
running Linux system, and \citet{rzepka2024causal} find causal inconsistencies to be normal in
Windows dumps as well. A causal inconsistency is a pair of structures in
one dump that could not have coexisted on the running system, because the
acquisition copied them at different moments: for example a doubly linked
kernel list whose forward and backward links no longer agree, a thread that
refers to a process absent from the process list, or a virtual address
descriptor tree that does not match the page tables it describes.
Part of what a rule layer flags on a dump may therefore have been
produced by the acquisition itself rather than by anything that happened on
the host.

\paragraph{Forensic datasets and ground truth}
A tool whose output is a judgement can only be tested against data
whose ground truth is known, and dataset availability and documentation remain
the weak point of the field~\citep{grajeda2017datasets}. Recent work asks how datasets should be
described, shared and built: \citet{breitinger2023sharing} propose a taxonomy
built on the organisation and the origin of the data, since origin decides
whether a dataset can be shared at all, and \citet{gobel2025datasets} set out
principles, characteristics and use cases for data set creation. Outside forensics, \citet{joyce2023motif}
measured what unverified labels cost: against expert ground truth, antivirus
majority voting assigns the correct family to $62.1\%$ of samples, and
the AVClass labeller to $46.8\%$. None of these studies looks
at the benign half of a paired memory corpus, which is where our problem
lies: whether the controls against which a detector is trained and scored
are in fact clean, and what a reported metric means when they are not.

\paragraph{Learning over memory artefacts}
Static analysis is limited by packing and obfuscation, and dynamic analysis
is imperfect in ways of its own, among them sandbox detection and
evasion, behaviour that depends on triggers or on time, the incomplete code
coverage of a single run and the dependence of the trace on the analysis
environment~\citep{ormeir2019dynamic}; memory sits between
the two.
DeepMem~\citep{song2018deepmem} learns graph neural models over a memory graph
to locate kernel objects. Its labels are structural, so ground truth can be
verified by construction, which is not true of compromise labels.
Feature-engineered detectors report near-perfect accuracy on
CIC-MalMem-2022~\citep{carrier2022malmem,dener2022malware},
whose 58{,}596 records are split evenly between benign and
malicious~\citep{carrier2022malmem}, and vision-based classifiers over process
dumps do similarly well~\citep{bozkir2021catch}. In their survey of
malware detection with graph representation learning,
\citet{bilot2024survey} conclude that feature design, label quality and
evaluation design matter more than the choice of classifier;
the memory-analysis survey of \citet{dehfouli2026memory} reviews memory
acquisition, the public corpora on which memory-based detectors are evaluated,
CIC-MalMem-2022 and the corpus of \citet{sihwail2021memory} among them, and the
detection methods built on them; and reviews of AI in digital forensics
chart the rapid growth of learned analysis~\citep{dunsin2024comprehensive}.
A different strand of work treats triage as sequential decision-making:
\citet{ghanem2026winregrl} plan over Registry and memory-timeline artefacts
with a Markov decision process and score the outcome at artefact level against
adjudicated ground truth, which is the kind of external adjudication that
snapshot-level labels lack. Message passing over graphs reaches the discriminative power of the
1-Weisfeiler--Lehman test under sum aggregation (GIN;~\citealp{xu2019gin});
GINE~\citep{hu2020pretraining} adds edge features to the message, which matters
when \code{injected\_into} and \code{belongs\_to} carry different forensic
meaning.

\paragraph{Evaluation validity and label quality}
\citet{apruzzese2023role} document a persistent discrepancy between
machine-learning research and practice in cybersecurity, and the base-rate
argument explains why a
plausible true-positive rate is of little operational use when the benign class
dominates~\citep{axelsson2000base}. \citet{arp2022dos} catalogue ten recurrent
pitfalls, including label inaccuracy, spurious correlations and inappropriate
baselines. Models that re-express their own curation rules are a case of
leakage~\citep{kapoor2023leakage}. Our work belongs to this line but moves the audit from the model to the
label.

\paragraph{Uncertainty and abstention}
Modern networks are systematically
miscalibrated~\citep{guo2017calibration,silvafilho2023calibration}. Monte-Carlo
dropout~\citep{gal2016dropout} and deep
ensembles~\citep{lakshminarayanan2017ensembles} provide variance-based proxies
for uncertainty, which themselves degrade under shift. One-class deep methods
offer a conformity view of ``normal'' behaviour~\citep{ruff2021unifying}. The reject option has
been surveyed by \citet{hendrickx2024reject}, who separate \emph{ambiguity}
rejection (the input lies near a decision boundary) from \emph{novelty}
rejection (the input lies outside what was seen in training); the two learned
paths of Section~\ref{sec:system} correspond to these two kinds. What is
rarely reported is the
abstention rate as a result in its own right. In forensic triage every
abstention is examiner time, so it has to be.

\paragraph{Gaps addressed by this paper} Read together, these strands
leave three gaps. First, the studies that report near-perfect memory-based
detection validate against labels asserted by provenance and do not examine
the benign class they rely on, and the dataset literature asks how data should
be described and shared but not how a benign baseline should be verified.
Second, the graph-learning and uncertainty literature supplies architectures,
calibration methods and reject options, but evaluates them on corpora whose
ground truth is taken as given, so a defect in the labels cannot be told from
a defect in the model. Third, the evaluation-validity literature audits models
and pipelines for leakage and pitfalls rather than the labels themselves, and
treats abstention as a design choice rather than as a measured cost. This
paper addresses the three in turn: it defines and measures label--signal
mismatch on the benign class (Section~\ref{sec:problem}), it builds an
architecture whose learned and rule-based parts can be separated and whose
deferrals are logged (Section~\ref{sec:system}), and it evaluates both with a
construct-validity audit and a reporting protocol that treat the abstention
rate, the fold composition and the label-quality fields as results
(Sections~\ref{sec:construct}--\ref{sec:discussion}).
%% ==================================================================
\section{Problem formulation: label--signal mismatch}
\label{sec:problem}
%% ==================================================================

This section defines the quantity the paper measures. It introduces the
mismatch rate and the indicator gap, names the three physical causes that can
produce them, distinguishes mismatch from classical label noise, and then
describes the governance schema that records label difficulty per row and
the scope of the claims that the pipeline makes.

Let $x$ be a memory snapshot with label $y\in\{0,1\}$ asserted from
provenance (here the capture folder: \code{WithVirus}${}\to 1$,
\code{NoVirus}${}\to 0$), and let $z\in\{0,1\}$ be the unobserved compromise
state of the host. \badd{Three objects are kept apart in what follows. The
\emph{provenance-assigned label} $y$ is observable and is what every public
corpus supplies. The \emph{host state} $z$ is latent; it becomes observable
only through an independent verification step (a rebuild from known-good
media, an allow-list check of every loaded image, an adjudicated manual
examination), and no corpus used in this paper records one. The \emph{rule
verdict} $r(x)$ is observable, reproducible by hand from plugin output, and
itself unvalidated: its error rates against $z$ are not known. Everything the
paper measures on the benign class is therefore a relation between $y$ and
$r(x)$, two observables, and not a measurement of $z$.}

\begin{definition}[Label--signal mismatch]
\label{def:mismatch}
Let $\mathcal{D}=\{(x_i,y_i)\}$ be a corpus with
$\mathcal{D}_c=\{x_i : y_i=c\}$, let
$r(x)\in\mathcal{V}=\{\textsc{low}\prec\textsc{medium}\prec\textsc{high}\prec\textsc{critical}\}$
be the verdict of a transparent rule layer and $\phi(x)\in\mathbb{R}^K$ the
indicators on which it rests. The \emph{mismatch rate} and the
\emph{indicator gap} are
\begin{align*}
\mu(\mathcal{D})&=\frac{|\{x\in\mathcal{D}_0 : r(x)\succeq\textsc{high}\}|}{|\mathcal{D}_0|},\\
\delta_k&=\bar\phi_k(\mathcal{D}_1)-\bar\phi_k(\mathcal{D}_0)
\end{align*}
$\mathcal{D}$ exhibits label--signal mismatch when $\mu$ is large; indicator
$k$ is \emph{inverted} when the rules treat it as monotone in risk yet
$\delta_k<0$.
\end{definition}

\rfix{The mismatch rate is a rate of \emph{rule--label discordance}:
$\mu=\Pr[r(x)\succeq\textsc{high}\mid y=0]$ estimated on $\mathcal{D}_0$. It
is not a contamination rate $\Pr[z=1\mid y=0]$, and the paper never reads it
as one. A large $\mu$ states that the benign class, as labelled, cannot be
distinguished from the malicious class by the indicators a practitioner would
cite to justify escalation, and therefore that the label ``benign'' is
unverified with respect to exactly those indicators.} It has three physical
causes, which a label alone cannot apportion.
\emph{Instrument noise}: rules fire on legitimate behaviour such as
security-product hooks or just-in-time
code~\citep{case2019hooktracer,ligh2014art}. \emph{Latent contamination}: a
baseline was never verified clean ($y=0$ but $z=1$). \emph{Acquisition
artefacts}: a non-atomic capture yields inconsistent
structures~\citep{pagani2019temporal,rzepka2024causal}.

Mismatch is not classical label noise~\citep{frenay2014label}, which assumes a
correct latent label corrupted by an independent process and is handled by
robust losses or relabelling. Here the latent state may be genuinely ambiguous.
A baseline host whose endpoint agent injects into \code{lsass.exe} is not
mislabelled as benign, and yet it is not clean in the sense the detector is
asked to learn. Formally, if $p_c(e)$ is the density of the observable evidence
$e$ under $z=c$ and $\pi=\Pr[z=1]$, no detector has risk below
$R^{\ast}=\int\min\{\pi p_1(e),(1-\pi)p_0(e)\}\,de$. As the two evidence
distributions overlap, $R^{\ast}\to\min\{\pi,1-\pi\}$ whatever the capacity of
the model, and any metric computed against $y$ measures agreement with
provenance, not with compromise.

\paragraph{Governance schema}
We therefore record label difficulty in the dataset manifest and not in the
discussion section. Each row carries four fields: \code{benign\_subtype} $\in$
\{\code{clean\_benign}, \code{hard\_benign\_admin\_tooling},
\code{ambiguous\_novirus\_control}\}; \code{label\_quality\_flag}, set when the
rule layer contradicts $y$ \rfix{under a fixed predicate on $r(x)$ and
$\phi(x)$ that is stated with the results (Section~\ref{sec:mismatch})};
\code{label\_quality\_reason}, the contradicting
indicators; and \code{train\_eligible}, whether the row may enter strict
training. The pipeline writes the fields, the manifest versions them, and the
training filters read them. An unfalsifiable claim (``our benign set is
clean'') becomes one that can be audited, and a reader can reconstruct which
rows supported which number. \rfix{Two consequences of the schema are easily
conflated and are kept apart here. A row with \code{train\_eligible=False} is
unsuitable for strict benign training because an instrument contradicts its
label; that is a statement about the evidence available for the row. It is
not a statement that the host was compromised, and the schema has no field
that would record one, because no rule-layer output can establish it.}

\badd{\paragraph{Extension for verified baselines} The four fields record
what a pipeline can observe. They cannot record what only the corpus builder
knows, and a benign class is verified, rather than asserted, only when those
fields exist. We therefore propose, and do not claim to have populated, the
following per-row extension for paired memory corpora: \code{label\_origin}
(folder, descriptor, analyst), \code{host\_id} and \code{capture\_id} (so that
pairs and repeated captures can be grouped), \code{acquisition} (tool,
version, guest paused or live, uptime at capture),
\code{resident\_security\_tools}, \code{known\_good\_attribution} (share of
loaded images and of executable private regions attributed to allow-listed
hashes), \code{verified\_state} $\in$ \{\code{clean}, \code{compromised},
\code{disputed}, \code{unverified}\} with \code{verification\_method} and the
evidence that supports a \code{clean} status, \code{rule\_label\_discordance}
(the present flag and reason), and \code{adjudication} (reviewer, date,
decision). The field \code{verified\_state} is the only one that speaks to
$z$, and it is filled by a verification step, never by the rule layer. Section~\ref{sec:discussion} says what each verification step costs and what
it excludes.}

\paragraph{Scope of claims}
The pipeline claims \emph{provenance traceability} (every score resolves to a
graph, plugin rows and a rule verdict), \emph{governance transparency},
\emph{recorded deferral} (the system may decline to decide, and logs why) and
\emph{falsifiable evaluation}.
It claims nothing about detection performance relative to production endpoint
tooling, about generalisation to unseen families, or about robustness to an
adaptive adversary~\citep{jin2021adversarial}; the threat model is explicitly
non-adaptive. Under this model the adversary is a malware sample that
executes on the guest and leaves whatever artefacts its execution leaves; it
is not assumed to know the rule layer, the graph schema or the learned models,
and it does not alter its behaviour to evade them. Hiding injected code from
VAD-based indicators~\citep{block2019pte}, poisoning the training corpus and
perturbing the graph~\citep{jin2021adversarial} are therefore outside the
evaluation, which is one reason the results are reported as a validation of
the corpus rather than of a detector. Every deferral is work passed to an examiner. A
system that defers often is expensive, not safe, and we report the deferral
rate as a cost.

\begin{table}[t]
\scriptsize
\caption{Memory-snapshot graph schema. Structural relations are read directly
from plugin output; intent relations are asserted by the rule layer and are
therefore the locus of the leakage risk discussed in Section~\ref{sec:system}.}
\label{tab:schema}
\scriptsize
\setlength{\tabcolsep}{4pt}
\begin{tabularx}{\columnwidth}{@{}l>{\raggedright\arraybackslash}X@{}}
\toprule
\textbf{Node types} & \code{process}, \code{thread}, \code{dll},
  \code{memory\_region}, \code{network\_conn}, \code{ip\_address},
  \code{handle}, \code{driver}, \code{kernel} \\
\midrule
\textbf{Structural edges} & \code{spawned\_by}, \code{belongs\_to},
  \code{loaded\_into}, \code{allocated\_in}, \code{connects\_from},
  \code{connects\_to} \\
\midrule
\textbf{Intent edges} & \code{injected\_into}, \code{intent\_injection},
  \code{intent\_c2}, \code{lolbin\_execution\_chain}, persistence and
  credential-access relations \\
\midrule
\textbf{Attribute profiles} & \code{full} $\cdot$
  \code{no\_manifest\_leakage} (rule-derived manifest fields masked) $\cdot$
  \code{structure\_only} (scale, edge distribution, motif tails) \\
\bottomrule
\end{tabularx}
\end{table}

%% ==================================================================
\section{Pipeline and models}
\label{sec:system}
%% ==================================================================

This section describes the pipeline stage by stage: the extraction of
plugin output from a memory image, its compilation into a typed graph, the
fixed rule layer, the two learned tracks (a calibrated binary classifier and
a pair of one-class conformity encoders), the fusion of their scores with
rule risk, and the abstention gate that decides whether a snapshot receives a
verdict or is routed to an examiner.

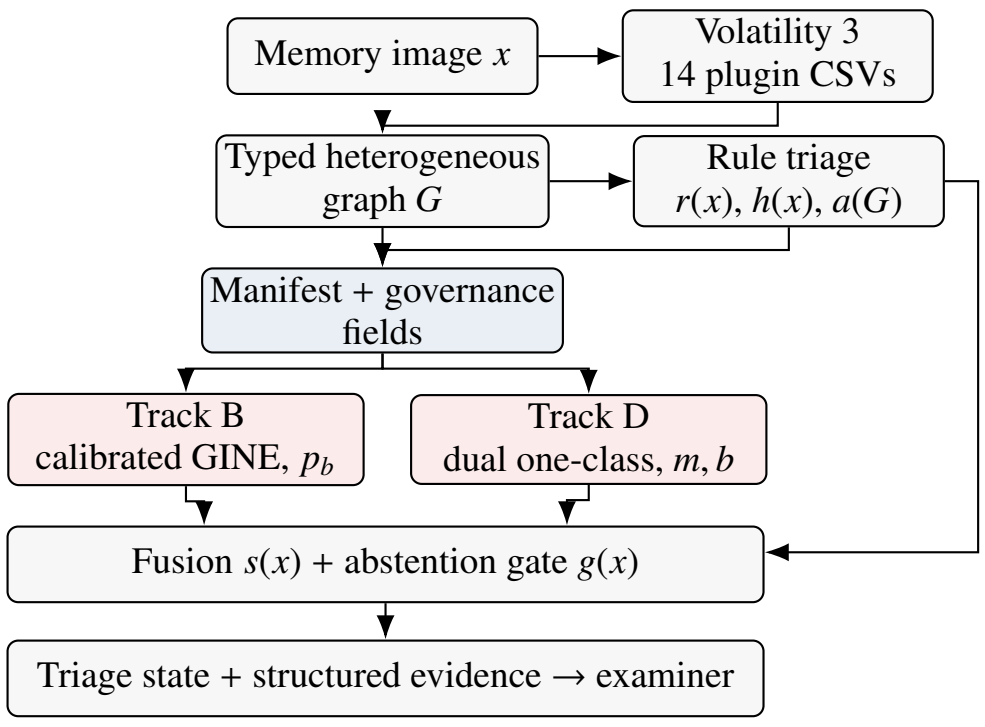
\begin{figure}[t]
\centering
\begin{tikzpicture}[
  font=\scriptsize,
  stage/.style={draw, rounded corners=2pt, align=center,
                minimum width=2.05cm, minimum height=0.50cm, inner sep=1.5pt,
                fill=black!3},
  gov/.style={draw, rounded corners=2pt, align=center, fill=benignblue!12,
              minimum width=2.05cm, minimum height=0.50cm, inner sep=1.5pt},
  ml/.style={draw, rounded corners=2pt, align=center, fill=malwarered!12,
             minimum width=2.05cm, minimum height=0.50cm, inner sep=1.5pt},
  ar/.style={-{Latex[length=1.6mm]}, thin}
]
\node[stage] (dump) {Memory image $x$};
\node[stage, right=0.55cm of dump] (vol) {Volatility 3\\14 plugin CSVs};
\node[stage, below=0.26cm of dump] (graph) {Typed heterogeneous\\graph $G$};
\node[stage, right=0.55cm of graph] (rules) {Rule triage\\$r(x)$, $h(x)$, $a(G)$};
\node[gov, below=0.26cm of graph] (man) {Manifest + governance\\fields};
\node[ml, below=0.26cm of man, xshift=-1.30cm, minimum width=2.35cm] (bin) {Track~B\\calibrated GINE, $p_b$};
\node[ml, right=0.30cm of bin, minimum width=2.35cm] (dual) {Track~D\\dual one-class, $m,b$};
\node[stage, below=0.26cm of bin, xshift=1.32cm, minimum width=5.00cm] (fuse) {Fusion $s(x)$ + abstention gate $g(x)$};
\node[stage, below=0.24cm of fuse, minimum width=5.00cm] (out) {Triage state $+$ structured evidence $\rightarrow$ examiner};

\draw[ar] (dump) -- (vol);
\draw[ar] (vol.south) -- ++(0,-0.15) -| (graph.north);
\draw[ar] (graph) -- (rules);
\draw[ar] (rules.south) -- ++(0,-0.15) -| (man.north);
\draw[ar] (graph.south) -- (man.north);
\draw[ar] (man.south) -- ++(0,-0.10) -| (bin.north);
\draw[ar] (man.south) -- ++(0,-0.10) -| (dual.north);
\draw[ar] (bin.south) -- ++(0,-0.10) -| ([xshift=-1.2cm]fuse.north);
\draw[ar] (dual.south) -- ++(0,-0.10) -| ([xshift=1.2cm]fuse.north);
\draw[ar] (rules.east) -- ++(0.22,0) |- ([yshift=0.08cm]fuse.east);
\draw[ar] (fuse) -- (out);
\end{tikzpicture}
\caption{Pipeline. Extraction and rule-based triage (grey) remain valid
independently of the learned components (red); the governance manifest
(blue) mediates every training and inference decision and is the artefact
that makes label difficulty auditable.}
\label{fig:pipeline}
\end{figure}

Figure~\ref{fig:pipeline} shows the flow and Algorithm~\ref{alg:triage} the
logic applied to one snapshot. Each stage writes an artefact that can be
inspected, so an examiner who distrusts a model score can fall back on the
triage JSON or on the originating plugin row without retraining anything.

\paragraph{Extraction}
Fourteen Volatility~3 plugins run per sample (\code{pslist}, \code{psscan},
\code{pstree}, \code{malfind}, \code{netscan}, \code{cmdline},
\code{dlllist}, \code{handles}, \code{threads}, \code{vadinfo},
\code{filescan}, \code{driverscan}, \code{ssdt},
\code{registry.hivelist}\rfix{, named here without their \code{windows.}
prefix};~\citealp{volatility3}) and write per-sample CSVs. \rfix{The plugin
set fixes a lower bound on the framework release: \code{windows.threads} was
introduced in Volatility~3 release 2.8.0 (October 2024), and in the current
releases \code{malfind} is the deprecated alias of
\code{windows.malware.malfind}. The release used by the pipeline, together
with the symbol-table (ISF) source for each guest, is recorded in the released
configuration.}\anote{insert the Volatility~3 release and the ISF source; state
whether any guest needed a manually supplied PDB or produced a plugin
failure.} Missing output sets a health flag instead of failing silently, so an incomplete
sample cannot enter training unnoticed. Three flags are written per
sample and gate every later stage: \code{filter\_ok} (all fourteen plugins
returned parseable output), \code{graph\_ok} (the graph compiled with at least
one process node and every referenced identifier resolved) and
\code{analyze\_ok} (rule triage and model scoring completed). Every row
reported in this paper carries all three (Section~\ref{sec:setup}). \badd{The
flags test that output was produced and could be parsed; they do not test that
the output is correct. A plugin that silently misses an artefact, as
\code{malfind} does for injected pages hidden from the VAD
view~\citep{block2019pte}, passes \code{filter\_ok}. Parseable output and
validated output are different things, and the pipeline certifies only the
first.}

\paragraph{Graph construction}
CSV rows are compiled into a typed heterogeneous graph $G=(V,E,\tau,\rho,X)$
with node-type map $\tau$, edge-type map $\rho$ and node features $X$
(Table~\ref{tab:schema}), loaded through PyTorch Geometric.
Edges divide into \emph{structural} relations $E_s$, read from plugin output
and reflecting Windows object relationships, and \emph{intent} relations $E_i$,
asserted by the rule layer and aligned with documented techniques such as
Process Injection (T1055) and Data Encrypted for Impact
(T1486).\footnote{MITRE ATT\&CK enterprise matrix,
\url{https://attack.mitre.org/matrices/enterprise/}, accessed 28 May 2026.}

\paragraph{From plugin rows to graph elements} Each plugin row becomes
one node or one edge, and every node keeps the identifiers of its source row
(plugin, process identifier, offset), which is what lets a score be traced
back to the output it came from. Process nodes are read from \code{pslist}
and \code{psscan}, with parentage from \code{pstree} written as
\code{spawned\_by}; a process present in \code{psscan} but absent from
\code{pslist} is retained as a process node and marked hidden. Threads,
handles and loaded modules become \code{thread}, \code{handle} and \code{dll}
nodes joined to their owning process by \code{belongs\_to} or
\code{loaded\_into}. Each \code{vadinfo} region and each \code{malfind} hit
becomes a \code{memory\_region} node joined by \code{allocated\_in} and
carrying the RWX, MZ-header, private-memory and commit-charge flags as
features. \code{netscan} rows become \code{network\_conn} nodes with
\code{connects\_from} to the owning process and \code{connects\_to} an
\code{ip\_address} node shared across connections; \code{driverscan} and
\code{ssdt} rows populate the \code{driver} nodes and the single
\code{kernel} node; \code{cmdline}, \code{filescan} and
\code{registry.hivelist} contribute node features and rule inputs rather than
nodes. Intent edges are added afterwards by the rule layer from these
features: \code{injected\_into} from \code{malfind} regions,
\code{intent\_c2} from connections that the rules class as
command-and-control style, \code{lolbin\_execution\_chain} from parent--child
pairs of living-off-the-land binaries, and the persistence and
credential-access relations in the same way. Because every intent edge is
derived from features that already sit on the nodes, the \code{structure\_only}
profile of Table~\ref{tab:schema} removes the intent edges together with the
rule-derived attributes.

\paragraph{Features and leakage control}
Node features combine type encodings with forensic indicators (RWX and
MZ-header flags, private memory, commit charge, load counts, per-node edge-role
counts). Graph-level attributes $a(G)$ summarise scale, density, injection and
C2 counts, attack-step and persistence counts, and motif tails. Many of these
come from the rule layer, and that creates an obvious circularity: a model fed
$r(x)$-derived attributes and evaluated on a corpus curated with $r(x)$ can
appear to learn while merely restating the rules. We therefore \rfix{define}
three attribute profiles\rfix{, which receive the following inputs}:
\code{full} \rfix{(all node features, all structural and intent edges, and the
complete $a(G)$)}; \code{no\_manifest\_leakage}, which masks
rule-derived manifest fields (maximum score, attack steps, injection and C2
counts, triage and subtype flags) \rfix{from $a(G)$ but keeps the node
features and the intent edges}; and \code{structure\_only} \rfix{(structural
edges only, node features reduced to type encodings and the plugin-level
counts that no rule interprets, and $a(G)$ reduced to scale, density and
degree-distribution summaries). The per-node edge-role counts and the
\code{memory\_region} nodes derived from \code{malfind} are themselves
instrument output, so even \code{structure\_only} is free of rule-layer
scoring but not of heuristic extraction. Every learned result in
Sections~\ref{sec:construct}--\ref{sec:gate} was produced under the
\code{full} profile; the other two profiles were specified but not trained or
scored in this study, and the leakage argument of
Section~\ref{sec:construct} rests on a comparison with a rule-attribute
baseline rather than on a profile ablation.}

\paragraph{Rule triage} The rule layer is a fixed, human-readable
scoring of the graph that runs before, and independently of, any learned
component.\anote{list the indicator weights and the four band thresholds in a
short table or in the released configuration; the verdicts in
Tables~\ref{tab:governance} and~\ref{tab:pairs_case} cannot be reproduced
by a reader without them} Each process accumulates points from the indicators attached to
it and to its regions, connections and children (injected regions, hidden
status, C2-style connections, living-off-the-land chains, persistence and
credential-access hits), and the snapshot verdict $r(x)$ is the band of the
maximum process score, from \textsc{low} to \textsc{critical}. The heuristic
risk $h(x)\in[0,1]$ used by the fusion layer is that maximum score rescaled to
the unit interval, and the graph-level attributes $a(G)$ carry the same
counts in aggregate; Table~\ref{tab:governance} reports three of them. The
layer is deliberately simple. Its verdicts can be reproduced by hand from the
plugin output, which is what qualifies it as a witness against the labels in
Definition~\ref{def:mismatch}; its own error characteristics are not assumed
to be known.

\paragraph{Track~B: calibrated binary path}
With $h_v^{(0)}=X_v$, each of $L$ GINE layers computes
\begin{equation}
\begin{aligned}
h_v^{(\ell)}&=\mathrm{MLP}^{(\ell)}\big((1+\epsilon^{(\ell)})\,h_v^{(\ell-1)}+m_v^{(\ell)}\big),\\
m_v^{(\ell)}&=\sum_{u\in\mathcal{N}(v)}\mathrm{ReLU}\big(h_u^{(\ell-1)}+e_{\rho(u,v)}\big),
\end{aligned}
\label{eq:gine}
\end{equation}
where $e_{\rho(u,v)}$ is a vector encoding of the edge
type~\citep{hu2020pretraining}. The graph representation concatenates mean, max
and sum readouts of $\{h_v^{(L)}\}$ with $a(G)$ and feeds a classification head
that produces logits $\mathbf{z}(x)$. Sum aggregation keeps the layer injective
on neighbour multisets~\citep{xu2019gin}, but on graphs where one node type
dominates, the readout is dominated as well; this is what the suspect-mask
pooling ablation of Section~\ref{sec:construct} probes. One forward pass
costs $O\big(L\,(|V|+|E|)\,d\big)$ for hidden width $d$, so even the largest
graph in the corpus (27{,}362 nodes, 39{,}117 edges) is cheap next to the
Volatility run that produces it. Training uses class-weighted loss and early
stopping on a held-out fold, followed by temperature
scaling~\citep{guo2017calibration}:
\begin{equation}
\begin{aligned}
p_b(x)&=\mathrm{softmax}\big(\mathbf{z}(x)/T\big)_{1},\\
T&=\arg\min\nolimits_{T'>0}\,\mathrm{NLL}_{\mathrm{val}}(T').
\end{aligned}
\label{eq:temp}
\end{equation}
Because $\mathbf{z}\mapsto\mathbf{z}/T$ is monotone, $T$ changes confidence but
not ranking. AUROC does not depend on it, and calibration therefore cannot
repair a ranking defect. The classifier is trained by minimising a
class-weighted cross-entropy and stopped early on the held-out fold; the strict split of Section~\ref{sec:setup} reserves 8 of the 43
snapshots for that purpose. The temperature is then fitted by a bounded
one-dimensional search on the validation negative log-likelihood whose upper
limit is $20$. Two variance estimates accompany every probability. The
Monte-Carlo dropout variance $\mathrm{Var}_{\mathrm{MC}}[p_b]$ is the variance
of $p_b$ over repeated forward passes with dropout kept active at
inference~\citep{gal2016dropout}; the ensemble variance
$\mathrm{Var}_{\mathrm{ens}}[p_b]$ is the variance of $p_b$ across
independently seeded training runs of the same
architecture~\citep{lakshminarayanan2017ensembles}. Both are compared with
fixed thresholds $\theta_v$ and $\theta_e$ in the abstention gate, and $p_b$
counts as saturated when it lies within a fixed margin of $0$ or $1$. The
number of layers $L$, the hidden width $d$, the dropout rate, the number of
stochastic passes, the ensemble size, the optimiser settings and the gate
thresholds are fixed before evaluation and recorded in the released
configuration.\anote{insert $L$, $d$, dropout rate, number of MC passes,
ensemble size, optimiser and learning rate, $\theta_v$, $\theta_e$,
$\theta_\Delta$ and the saturation margin; the manuscript cannot be reproduced
from the text while these remain symbolic.} \rfix{The eight validation graphs
of the strict split serve both early stopping and the temperature fit, so the
calibration figures reported on them in Section~\ref{sec:binary} are not
held out with respect to $T$.} The analysis stage does not threshold at $0.5$. It
assigns four bands using $p_b$ and a narrow ambiguity interval
$[\tau_{\mathrm{low}},\tau_{\mathrm{high}}]$: \state{likely benign},
\state{low-risk ambiguous}, \state{high-risk ambiguous} and
\state{likely malicious}, where the two ambiguous states are separated by
heuristic risk $h(x)$ and not by $p_b$.

\paragraph{Track~D: dual conformity path}
Two one-class GINE encoders are trained
separately~\citep{ruff2021unifying}: a malware-pattern model
yielding $m(x)\in[0,1]$ and a benign-conformity model yielding $b(x)\in[0,1]$,
with margin $\Delta(x)=m(x)-b(x)$. Each encoder has the layer structure of
Eq.~\eqref{eq:gine} followed by a graph readout, and is trained on the
snapshots of its own class only under a one-class objective that draws the
embeddings of that class towards a common centre~\citep{ruff2021unifying}. At
inference the distance of a snapshot's embedding to the centre is mapped
monotonically to the unit interval, so that $m(x)$ is high when the snapshot
resembles the malware-labelled training graphs and $b(x)$ is high when it
resembles the benign-labelled ones\rfix{; writing $\delta_c(x)$ for the
distance to the centre of class $c$, the map is $m(x)=\psi(\delta_{\mathrm{mal}}(x))$
and $b(x)=\psi(\delta_{\mathrm{ben}}(x))$ with $\psi:[0,\infty)\to[0,1]$
decreasing and fixed before evaluation}\anote{insert the exact form of $\psi$
and of $d(m,b)$ from the code}. The dual signal $d(m,b)\in[0,1]$ is a fixed
function of the margin $\Delta=m-b$, increasing in $\Delta$, \rfix{and all
three terms of the fusion below lie in $[0,1]$ and are oriented so that larger
values mean greater risk; since the weights sum to one, $s(x)\in[0,1]$}. A fusion layer forms the convex combination
\begin{equation}
s(x)=w_b\,p_b(x)+w_d\,d\big(m(x),b(x)\big)+w_h\,h(x),
\label{eq:fusion}
\end{equation}
with $(w_b,w_d,w_h)=(0.35,0.40,0.25)$, where $d$ is the dual-model signal,
thresholded into triage states at $(0.40,0.65)$. The intention is that a
snapshot may be suspicious without being recognised:
\state{anomalous unknown} is a legitimate output, distinct from both classes\rfix{,
and is assigned before any gate when neither encoder recognises the snapshot,
that is when $m(x)$ and $b(x)$ both fall below a fixed floor}\anote{confirm the
condition and its threshold; two snapshots receive this state under the
\code{disabled} preset in Table~\ref{tab:dual}, one of them the
malware-labelled snapshot at $(b,m)=(0.000,0.011)$}.
\begin{algorithm}[t]
\caption{Abstention-first triage of one memory image}
\label{alg:triage}
\begin{algorithmic}[1]
\Require memory image $x$; gate preset $\pi$; fusion weights $(w_b,w_d,w_h)$;
  triage band $[\tau_1,\tau_2)=[0.40,0.65)$; gate thresholds
  $\theta_v,\theta_e,\theta_\Delta$
\Ensure triage state $q$, fused score $s$, trigger set $\mathcal{T}$, evidence
  pointers
\State Run the fourteen Volatility~3 plugins on $x$, compile the typed graph
  $G$ and set the health flags
\State Apply the rule layer to $G$ to obtain the verdict $r$, the heuristic
  risk $h$ and the graph attributes $a$
\State Write the governance fields of Section~\ref{sec:problem} to the manifest
\State $p_b \gets \mathrm{softmax}\big(f_{\theta}(G,a)/T\big)_1$
  \Comment{Track~B, Eq.~\eqref{eq:temp}}
\State $m \gets f_{\mathrm{mal}}(G)$,\; $b \gets f_{\mathrm{ben}}(G)$,\;
  $\Delta \gets m-b$ \Comment{Track~D}
\State $s \gets w_b\,p_b + w_d\,d(m,b) + w_h\,h$ \Comment{Eq.~\eqref{eq:fusion}}
\If{$s \ge \tau_2$}
  \State $q \gets$ \state{likely malicious}
\ElsIf{$s < \tau_1$}
  \State $q \gets$ \state{likely benign}
\Else
  \State $q \gets$ \state{needs analyst review} \Comment{or \state{anomalous unknown}, see text}
\EndIf
\State $\mathcal{T} \gets \emptyset$
\If{$\mathrm{Var}_{\mathrm{MC}}[p_b] > \theta_v$} add \textit{dropout variance} to $\mathcal{T}$ \EndIf
\If{$\mathrm{Var}_{\mathrm{ens}}[p_b] > \theta_e$} add \textit{ensemble variance} to $\mathcal{T}$ \EndIf
\If{$|\Delta| < \theta_\Delta$} add \textit{low margin} to $\mathcal{T}$ \EndIf
\If{$p_b$ and $(m,b)$ disagree \textbf{and} $p_b$ is not saturated}
  add \textit{disagreement} to $\mathcal{T}$
\EndIf
\If{the gate of preset $\pi$ fires on $\mathcal{T}$}
  \State $q \gets$ \state{needs analyst review}; log $\mathcal{T}$
\EndIf
\State \Return $q$, $s$, $\mathcal{T}$ and pointers to $G$, the plugin rows and
  the rule hits
\end{algorithmic}
\end{algorithm}
\paragraph{Abstention gate}
An uncertainty gate can override any decisive state with
\state{needs analyst review} and records the trigger: high Monte-Carlo dropout
variance~\citep{gal2016dropout}, high ensemble
variance~\citep{lakshminarayanan2017ensembles}, low dual margin, or
binary--dual disagreement. The last is suppressed when $p_b$ is saturated,
since an overconfident head is not evidence of uncertainty. Three presets
make the routing behaviour itself something that can be ablated
(Section~\ref{sec:gate}): \code{disabled} applies no gate, so a snapshot
is routed by the fused score alone; \code{legacy\_or} fires on any of the four
triggers, including disagreement whether or not $p_b$ is saturated; and
\code{calibrated}, the preset behind every routing result reported here,
applies the trigger set $\mathcal{T}$ of Algorithm~\ref{alg:triage} as written,
with the saturation exception. In selective-classification
terms~\citep{hendrickx2024reject}, the gate is a selection function
$g(x)=\mathbf{1}[q(x)\ \text{decisive}]$ with coverage
$\varphi=\mathbb{E}[g(x)]$ and selective risk
$R=\mathbb{E}[\ell(\hat y,y)\,g(x)]/\varphi$. A reject rule based on the
posterior is optimal only when the posterior can be trusted. Where it cannot,
the honest summary of an
abstaining system is the pair $(\varphi,R)$ together with an ablation showing
what produces $\varphi$.

%% ==================================================================
\section{Experimental setup}
\label{sec:setup}
%% ==================================================================

This section describes the corpus and its two manifest views, the
grouping and splitting protocol, the construct-validity audit step by step,
and the metrics and derived statistics reported in
Sections~\ref{sec:mismatch}--\ref{sec:gate}.

\paragraph{Corpora}
Samples derive from MalVol-25, a public Windows volatile-memory dataset of
paired clean and infected virtual-machine
snapshots~\citep{dunsin2025malvol,dunsin2025malvoldata}. We use two manifest views over the
same artefact contracts. The descriptive corpus ($n{=}30$) comprises
fifteen families, each contributing one infected capture and one control from
the same host configuration: ten ransomware families (Cerber, DeriaLock,
Dharma, GandCrab, GoldenEye, InfinityCrypt, Locky.AZ, LuckyLocker, Spora,
WannaCry), two worms (MyDoom, BlackWorm), PowerLoader, RedTail (a cryptominer)
and a DLL-hijacking scenario, categorised as in the dataset descriptor, across
Windows~7, 8.1, 10 and~11 guests. The ML corpus ($n{=}43$; 25
malware-labelled, 18 benign-labelled) retains the paired rows, adds further
captures from the same lineage, and is the authoritative source for all learned
results. All reported rows passed the \code{filter\_ok}, \code{graph\_ok} and
\code{analyze\_ok} health checks. The descriptor states that each clean
snapshot was taken from the guest in its uninfected state and validated by
Volatility extraction and manual inspection. It records no independent check
that baselines are free of injection-like activity. That is typical of the
field and not a peculiarity of this dataset, and it is the reason we write
\emph{benign-labelled} throughout. The descriptor reports clear differences
between clean and infected states under manual comparison; our question is
whether those differences survive translation into the thresholded indicators
that automated triage depends on. \badd{The descriptor does not state the
acquisition tool, whether the guest was paused or running when its memory was
read, or the uptime at capture, so the acquisition-artefact cause of
Section~\ref{sec:problem} can be neither confirmed nor excluded for any
row.}\anote{state the origin of the 15 ML-corpus captures that are not members
of the 14 complete pairs (11 malware-labelled, 4 benign-labelled): which
families and hosts they belong to decides whether holding out a pair also
holds out its family} Two further public corpora, used for
external validation of the protocol rather than of the pipeline, are described
in Section~\ref{sec:ext_setup}.

\paragraph{Protocol}
Both members of a family pair share a host baseline, so random splitting leaks
badly. Family is the grouping key throughout, and the construct-validity audit
uses strict LOFO over the 14 complete pairs of the ML corpus. This is a matched
instance of leave-pair-out cross-validation, the preferred AUROC estimator on
small samples~\citep{airola2011lpo}. Tabular and graph-attribute rungs refit a
logistic model per fold\rfix{, so their LOFO values are out-of-sample
estimates}. GNN rungs score a frozen checkpoint with softmax
$P(\text{malware})$ at $T{=}1$ and no per-fold refit\rfix{. Their LOFO values
are therefore in-sample ranking diagnostics of a checkpoint that has seen
every fold, and we report them as such; they are not out-of-sample
performance estimates, and we no longer describe them as upper bounds, since a
model trained on a fold is expected but not guaranteed to rank it at least as
well as one that was not}. The label-permutation null
uses $B{=}500$ permutations\rfix{, applied to the same frozen checkpoint without
retraining, so it tests whether the checkpoint's fixed ordering is associated
with the labels and not whether a model trained under the null would reach
the observed mean}. Because no benign-labelled row is train-eligible
(Section~\ref{sec:mismatch}), a strict trainer has no benign class to fit. Both
learned paths were necessarily fitted with the flagged controls standing in for
that class, and every learned result inherits this caveat\rfix{: the benign
one-class encoder of Track~D is, in effect, a contaminated-control experiment,
and Section~\ref{sec:dual} reads it as one}.

\paragraph{The construct-validity audit in full} The audit runs as
follows. (i)~For each of the $F{=}14$ complete family pairs the pair is held
out and the rung under test is scored on its two graphs; tabular rungs are
refitted on the remaining 41 snapshots\anote{confirm that the 41 exclude
every capture of the held-out family, not only the two members of the pair;
if other captures of that family remain in training, the tabular rungs are
leave-one-pair-out rather than leave-one-family-out}, GNN rungs use the frozen checkpoint.
The fold AUROC $A_f$ takes the value $1$, $\tfrac12$ or $0$
(Proposition~\ref{prop:twograph}), and the ladder reports the fold mean, the
population standard deviation across folds and the composition $(W,T,L)$.
(ii)~The label-permutation null re-labels the 43 snapshots at random
$B{=}500$ times, repeats step~(i) for the GNN rung, and reports the fraction of
permuted means at or above the observed mean as $p$, with Monte-Carlo standard
error $\sqrt{p(1-p)/B}$; Section~\ref{sec:construct} shows that this
permutation was not confined to pairs and gives the exact within-pair null
that should replace it.\anote{state how a fold whose two graphs received the
same permuted label was handled (dropped, scored $\tfrac12$, or re-drawn);
the result depends on it} (iii)~The paired diagnostics of Table~\ref{tab:pairs}
use the same 14 folds and the within-pair score difference $\Delta P_f$ with
tolerance $\eps{=}0.05$. (iv)~The suspect-mask ablation repeats step~(i) with
the readout restricted to nodes that carry at least one rule hit.
\badd{(v)~Because the GNN's ties are produced by saturation rather than by
chance, every sign test on a rung with ties is reported twice, with ties
dropped (the conventional test) and with ties counted as losses (a tie is a
failure to order the pair); the two agree when $T{=}0$.} Every
threshold that enters a reported number is fixed: the ambiguity interval
$[0.632,0.712]$ of Track~B, the fusion weights $(0.35,0.40,0.25)$ and triage
band $[0.40,0.65)$, the tolerance $\eps$, $B$, and the strict split of 35
training and 8 validation snapshots. \badd{No threshold was tuned on a test
fold; the temperature of Track~B is the one quantity fitted on held-out data,
and the eight graphs it was fitted on are also the eight on which its
calibration is reported (Section~\ref{sec:binary}).}

\paragraph{Metrics and derived statistics}
We report AUROC, the probability that a positive outranks a negative with ties
counted one half, together with the fold standard deviation; precision, recall,
F1 and \rfix{selective} accuracy on decisive predictions \badd{(the overall
accuracy that counts an abstention as an error is given alongside it)}; Brier score, expected calibration
error (ECE;~\citealp{guo2017calibration,silvafilho2023calibration}) and fitted
temperature; and, as results in their own right, review-routing rate, decisive
coverage and gate-trigger counts. Every proportion carries an exact 95\%
Clopper--Pearson interval~\citep{clopper1934}, paired comparisons use exact
one-sided sign tests with ties dropped, and the Monte-Carlo error of the
permutation $p$-value is stated~\citep{ojala2010permutation}. The Brier score is the mean squared difference between $p_b$ and the
manifest label; ECE partitions $[0,1]$ into equal-width bins and sums, over
bins, the bin's share of samples times the absolute gap between its mean
probability and its empirical accuracy; the Clopper--Pearson interval for $k$
successes in $n$ trials is the exact binomial interval whose end points are
the $\alpha/2$ quantile of $\mathrm{Beta}(k,n-k+1)$ and the $1-\alpha/2$
quantile of $\mathrm{Beta}(k+1,n-k)$; the one-sided sign test on $W$ wins and
$L$ losses reports $\Pr[X\ge W]$ for $X\sim\mathrm{Bin}(W+L,\tfrac12)$; and
pooled AUROC comparisons use the exact one-sided Mann--Whitney test. These derived
quantities are computed from the counts and per-sample values reported here by
a script that accompanies the artefacts. All primary numbers come from the
archived analysis files and the construct-validity report. The
derived-statistics and external-audit scripts were drafted with the
assistance of a generative-AI coding tool and were reviewed, run and checked
by the authors; every number they produce is reproducible from the archived
files.

%% ==================================================================
\section{Results}
%% ==================================================================

Results are reported in the order of the audit. Section~\ref{sec:mismatch}
examines the benign class itself, Section~\ref{sec:construct} asks whether the
learned model adds anything to the rules, Sections~\ref{sec:binary}
and~\ref{sec:dual} report the calibrated and the conformity path, and
Section~\ref{sec:gate} ablates the abstention gate.

\subsection{The benign class does not behave like a \rfix{verified} benign class}
\label{sec:mismatch}

\begin{table}[t]

\caption{Governance audit and indicator gaps. Every benign-labelled row in
both views is flagged and none qualifies as \code{clean\_benign}; two of three
headline indicators are inverted (Definition~\ref{def:mismatch}). \rfix{The
descriptive view records the flag and the ML view the eligibility field; a
dash marks a field the view does not record.}}
\label{tab:governance}

\setlength{\tabcolsep}{4pt}
\begin{tabular}{@{}lcc@{}}
\toprule
& \textbf{Descriptive} & \textbf{ML} \\
& ($n{=}30$) & ($n{=}43$) \\
\midrule
Malware- / benign-labelled rows   & 15 / 15 & 25 / 18 \\
{ambiguous\_novirus\_control} & 15/15 & 18/18 \\
{label\_quality\_flag} set    & 15/15 & --- \\
{train\_eligible=False}       & ---   & 18/18 \\
{clean\_benign}               & 0     & 0 \\
\bottomrule
\end{tabular}\\[5pt]
\begin{tabular}{@{}lrrr@{}}
\toprule
\textbf{Descriptive corpus} & \textbf{Benign} & \textbf{Malware} &
\textbf{Gap $\delta_k$} \\
\midrule
Injections (mean)            & 11.73 & 10.93 & $-0.80$ \\
C2-style connections (mean)  & 13.20 & 11.27 & $-1.93$ \\
Max.\ process score (mean)   & 76.60 & 79.80 & $+3.20$ \\
Nodes (median)               & 17{,}595 & 20{,}077 & \\
Edges (median)               & 25{,}684 & 25{,}635 & \\
\bottomrule
\end{tabular}
\end{table}

Table~\ref{tab:governance} states the central finding. In neither manifest view
does a single benign-labelled snapshot survive as a \rfix{verified} baseline. All 15
descriptive and all 18 ML benign rows are tagged
\code{ambiguous\_novirus\_control} (exact 95\% lower bounds on the flagged
share: $0.78$ and $0.82$), and on the ML corpus none is train-eligible under
the strict filter. The recorded reasons are ordinary forensic ones: HIGH or
CRITICAL verdicts, elevated maximum process scores, non-zero injection counts,
C2-style connections. \rfix{The flag is set when any one of these recorded
reasons is present, so the flag rate (15/15) and the mismatch rate $\mu$,
which counts verdicts of HIGH or above only, are different quantities and are
reported separately below.}\anote{confirm the flag predicate; the MyDoom
control of Table~\ref{tab:pairs_case} is flagged with a LOW verdict and no
injections, so state its recorded \code{label\_quality\_reason}; if the
\code{ambiguous\_novirus\_control} subtype was assigned to every control by
policy rather than by the predicate, say so, because the 15/15 would then be a
design decision and not a finding}

It is the size of the effect that surprised us. On the descriptive corpus the
rule layer returns HIGH or above on 25 of 30 samples, MEDIUM on 3 and LOW on 2.
Both LOW rows belong to one pair, so at least 11 of the 15 controls are rated
HIGH or CRITICAL and $\mu\ge0.73$\rfix{; the three MEDIUM verdicts bound it
from above at $14/15$}.\anote{the manifest records a verdict per row, so the
exact $\mu$ (11, 12, 13 or 14 of 15) is available and should replace the
bounds} The class-conditional means run the
wrong way for two of the three headline indicators. Benign-labelled rows
average 11.73 injections and 13.20 C2-style connections, against 10.93 and
11.27 for malware-labelled rows, and only maximum process score is marginally
higher on the malicious side (79.80 vs 76.60). A practitioner escalating on
injection or C2 volume alone would escalate the controls slightly more often
than the infected captures. Graph scale does not separate the classes either:
node counts run from 86 to 27{,}362 and edge counts from 7 to 39{,}117, both
extremes are occupied by both classes, and the medians differ by
2{,}482 nodes (17{,}595 against 20{,}077) and by 49 edges (25{,}684 against
25{,}635) (Table~\ref{tab:governance}).

\begin{table}[t]

\caption{Four family pairs from the descriptive corpus. In every pair the
control carries indicators that would justify escalation on its own; in two
of the four it is not clearly quieter than its infected partner.}
\label{tab:pairs_case}

\setlength{\tabcolsep}{3.7pt}
\begin{tabular}{@{}llrrl@{}}
\toprule
\textbf{Family} & \textbf{Capture} & \textbf{Nodes} & \textbf{Inj.} &
\textbf{Verdict} \\
\midrule
\multirow{2}{*}{GoldenEye}
 & infected & n/r & 41 & CRITICAL \\
 & control  & n/r &  8 & CRITICAL \\
\midrule
\multirow{2}{*}{Cerber}
 & infected & 7{,}904 & 6 & HIGH \\
 & control  & 9{,}382 & 6 & HIGH \\
\midrule
\multirow{2}{*}{WannaCry}
 & infected & 8{,}098 & 7 & HIGH \\
 & control  & 7{,}590 & 9 & HIGH \\
\midrule
\multirow{2}{*}{MyDoom}
 & infected &  164 & 0 & LOW \\
 & control  &   86 & 0 & LOW \\
\bottomrule
\end{tabular}
\\[2pt]
{\footnotesize Every control row is tagged
\code{ambiguous\_novirus\_control} with \code{label\_quality\_flag=True}.
The descriptive manifest records C2-style connections for the Cerber
pair only (24 infected, 7 control) and no node counts for the GoldenEye pair
(n/r); GoldenEye's control additionally reports 13 hidden processes.}
\end{table}

Individual pairs make the mechanism concrete (Table~\ref{tab:pairs_case}).
GoldenEye's infected capture records the corpus maximum of 41 injections, but
its own control records 8 injections, 13 hidden processes and a CRITICAL
verdict. Cerber's pair is comparable in scale (7{,}904 vs 9{,}382 nodes); both
members carry six injections and HIGH verdicts and differ mainly in C2 count
(24 vs 7). WannaCry's control reports more injections than its infected
counterpart (9 vs 7). Only the MyDoom pair, at 86 and 164 nodes far smaller
than the rest, returns LOW on both members, and even there the control is
flagged.

\subsection{Construct validity: does the model add anything?}
\label{sec:construct}

\begin{table}[t]
\caption{LOFO stage ladder over the 14 complete family pairs. W/T/L is the
fold composition implied by Proposition~\ref{prop:twograph}; sign-test
$p$-values are exact, one-sided, with ties dropped. \rfix{The two logistic
rungs are refitted per fold (out-of-sample); the two GNN rungs score a frozen
checkpoint trained on every fold (in-sample diagnostics).}}
\label{tab:ladder}

\setlength{\tabcolsep}{2.2pt}
\begin{tabular}{@{}lccccc@{}}
\toprule
\textbf{Stage} & \textbf{AUROC} & \textbf{SD} & \textbf{W/T/L} &
\textbf{Sign $p$} & \textbf{Null $p$} \\
\midrule
Manifest tabular (LR)        & 0.643 & 0.479 & 9/0/5  & 0.212 & --- \\
Graph attributes (LR)   & 0.786 & 0.410 & 11/0/3 & 0.029 & --- \\
GNN, global pooling\rfix{$^{\dagger}$}  & 0.786 & 0.311 & 9/4/1  & 0.011\rfix{$^{\ddagger}$} & 0.044 \\
GNN, suspect-mask\rfix{$^{\dagger}$}   & 0.786 & 0.311 & 9/4/1  & 0.011\rfix{$^{\ddagger}$} & --- \\
\midrule
\multicolumn{6}{@{}l}{\emph{Pooled AUROC, frozen checkpoint, all 43 snapshots}} \\
\multicolumn{6}{@{}l}{\quad Global $0.615$ \quad Suspect-mask $0.615$} \\
\multicolumn{6}{@{}l}{\quad $\Delta$ LOFO (mask $-$ global) $0.000$} \\
\multicolumn{6}{@{}l}{\rfix{\footnotesize $^{\dagger}$frozen checkpoint, in-sample; across-pair permutation null}} \\
\multicolumn{6}{@{}l}{\rfix{\footnotesize $^{\ddagger}$ties dropped; $p=0.212$ with the four ties counted as losses}} \\
\bottomrule
\end{tabular}
\end{table}

Table~\ref{tab:ladder} is the most useful diagnostic in the paper. Read
downwards it is encouraging: adding graph attributes lifts LOFO AUROC from
$0.643$ to $0.786$, and the learned model is separable from a label-permutation
null at $p{=}0.044$ (observed $0.786$ against $0.498\pm0.161$). Read across, it
deflates. The graph neural network matches, to three decimal places, a logistic
regression over rule-derived graph attributes, and restricting pooling to
suspect-flagged nodes changes nothing ($\Delta{=}0.000$; identical pooled AUROC
of $0.615$). \rfix{Read with the footnote, it deflates further. The two
logistic rungs are out-of-sample estimates; the two GNN rows are in-sample
diagnostics of a checkpoint that had seen both graphs of every fold. A
checkpoint that has seen a pair and still cannot order it by more than a
rounding margin is informative about the representation, not about
generalisation, and nothing in Table~\ref{tab:ladder} is evidence that the
network would reach $0.786$ on a family it had not seen.}

The standard deviations carry more information than is usually taken from them.

\begin{proposition}[Two-graph folds]
\label{prop:twograph}
Let each of $F$ folds hold one positive and one negative graph with scores
$s_f^{+},s_f^{-}$. Then the fold AUROC is
$A_f=\mathbf{1}[s_f^{+}>s_f^{-}]+\tfrac12\mathbf{1}[s_f^{+}=s_f^{-}]\in\{0,\tfrac12,1\}$.
If $W$, $T$ and $L$ folds take the values $1$, $\tfrac12$ and $0$, the fold
mean $\bar A$ and population variance $\sigma^2$ satisfy
\begin{equation}
T=4F\big(\bar A-\bar A^{2}-\sigma^{2}\big),\quad
W=F\bar A-\tfrac{T}{2},\quad L=F-W-T,
\label{eq:wtl}
\end{equation}
so $(\bar A,\sigma)$ determine $(W,T,L)$ uniquely.
\end{proposition}
\begin{proof}
$\bar A=(W+T/2)/F$ and $\mathbb{E}[A^2]=(W+T/4)/F$, hence
$\sigma^2=\bar A-T/(4F)-\bar A^2$; solve for $T$, then $W$ and $L$.
\end{proof}

Two things follow. First, $\bar A$ over matched two-graph folds is simply the
fraction of pairs ordered correctly, with ties counted one half. It sees only
within-pair order, and ``$0.786\pm0.311$'' summarises fourteen ternary
outcomes, not a smooth performance estimate. Second, Eq.~\eqref{eq:wtl}
recovers the composition from the published summary (Table~\ref{tab:ladder};
Figure~\ref{fig:scores}c). For the GNN row,
$T=56\,(0.786-0.786^{2}-0.311^{2})=4.0$, hence $W=11-2=9$ and $L=1$. \badd{The
reconstruction tolerates the rounding: for $F{=}14$ the 120 possible
compositions remain distinct after $\bar A$ and $\sigma$ are rounded to three
decimals, the nearest two compositions that share a mean differ in $\sigma$ by
$0.037$, and a non-integer $T$ from Eq.~\eqref{eq:wtl} is the sign that a
reported deviation is a sample rather than a population one (for the GNN row
the sample deviation would be $0.323$).} The two
$0.786$ rows turn out to be different objects. The attribute-only regression is
decisive in every fold and right in 11 of 14 (sign test $p{=}0.029$). The GNN
is right in 9, wrong in 1 and exactly tied in 4 ($p{=}0.011$ on ten
untied folds). Exact ties are what saturated outputs produce: the fitted
temperature of Section~\ref{sec:binary} implies logits an order of magnitude
too large, so at $T{=}1$ both members of a pair can round to the same
probability. \rfix{A saturated tie is not a chance event that the sign test
may discard; it is a failure to order the pair. Counted as losses, the four
ties give 9 wins against 5 non-wins and $p{=}0.212$, the same value as the
manifest-tabular rung. The significance of the GNN row therefore rests
entirely on the convention of dropping ties, and we report both values.} The
manifest-tabular rung (9/0/5; $p{=}0.21$) cannot be told from
guessing. The permutation result is fragile as well. With $B{=}500$ the
Monte-Carlo standard error of $p{=}0.044$ is $0.009$, and its 95\% interval
$[0.026,0.062]$ straddles $0.05$\rfix{; and because the checkpoint was scored,
not retrained, under each permutation, the $p$-value tests the association
between a fixed in-sample ordering and the labels, which is the weakest claim
a permutation test can support}.

\paragraph{What fourteen pairs can detect}
Proposition~\ref{prop:twograph} also fixes the resolution of the design. With
$F{=}14$ untied pairs, a one-sided sign test at $\alpha{=}0.05$ needs at least
11 wins, that is a LOFO mean of at least $11/14=0.786$. The headline value of
both models is therefore the smallest \rfix{mean that a fully decided design
of fourteen pairs can declare significant}, and the step from $0.643$ to
$0.786$ is two pairs. \rfix{The qualification matters: when ties are dropped,
a rung with many ties can be declared significant at a lower mean (seven wins
and seven ties give $\bar A=0.750$ and $p{=}0.008$), which is a further reason
to report the composition and not the mean alone.} Power
is correspondingly low. If a model truly ordered $75\%$ of matched pairs
correctly, fourteen pairs would detect it with probability $0.52$;
$80\%$ power at that effect needs 28 pairs, and 42 if the true rate is $70\%$
\badd{(the exact test's power is not monotone in $F$ because the critical
count moves in steps: 23 pairs first reach $0.80$ at a true rate of $75\%$
but 24 and 25 fall below it, and 28 is the smallest $F$ from which power
stays at or above $0.80$; the same convention gives 42 at $70\%$, where 37
pairs first reach it)}.
A paired corpus meant to support claims about learned triage should be sized
from such a calculation, not from the number of malware families that happen to
be available.

\begin{corollary}[Exact within-pair null]
\label{cor:null}
If labels are exchangeable within each pair, then $\bar A=(X+T/2)/F$ with
$X\sim\mathrm{Bin}(F-T,\tfrac12)$, the exact $p$-value is $\Pr[X\ge W]$, and
the null standard deviation of $\bar A$ is $\sqrt{F-T}/(2F)\le 1/(2\sqrt{F})$.
\end{corollary}

\noindent
The null has only $2^{14}=16{,}384$ configurations, so it can be enumerated and
the Monte-Carlo error above is avoidable; its $p$-values are the sign-test
values of Table~\ref{tab:ladder}. For $F{=}14$ the bound equals $0.134$,
whereas the reported permutation null has a standard deviation of $0.161$.
Since $0.161$ exceeds a bound that holds for every within-pair null,
the reported null was not a within-pair one: labels were permuted across
pairs, which breaks the matching and lets between-family variance back in
\badd{(under a permutation of all 43 labels a fold keeps one graph of each
class with probability $0.498$ only, so half the folds of a permuted replicate
have no defined AUROC)}.
The within-pair values are the ones to quote\rfix{, with the proviso of the
previous paragraph: for a frozen checkpoint they remain in-sample, and the
test that would support a generalisation claim, refitting the network without
the held-out family and enumerating the within-pair null on the refitted
scores, was not run in this study}.

\begin{table}[t]
\caption{Paired-control diagnostics over 14 family pairs with tolerance
$\eps{=}0.05$, where $\Delta P_f=P_{\mathrm{WV}}-P_{\mathrm{NV}}$ is the
within-pair score difference \rfix{of the frozen GNN checkpoint (in-sample)}.
Intervals are exact 95\% Clopper--Pearson.}
\label{tab:pairs}

\setlength{\tabcolsep}{3.4pt}
\begin{tabular}{@{}lccc@{}}
\toprule
\textbf{Diagnostic} & \textbf{Value} & \textbf{Pairs} & \textbf{95\% CI} \\
\midrule
Pair-rank accuracy ($\Delta P_f\ge\eps$)    & 0.286 & 4/14  & 0.08--0.58 \\
Pair inversion rate ($\Delta P_f\le-\eps$)  & 0.000 & 0/14  & 0.00--0.23 \\
Near-tie rate ($|\Delta P_f|<\eps$)         & 0.714 & 10/14 & 0.42--0.92 \\
\midrule
Zone suspect-count inversion                & 0.429 & 6/14  & 0.18--0.71 \\
Zone max-score inversion                    & 0.571 & 8/14  & 0.29--0.82 \\
\bottomrule
\end{tabular}
\end{table}

Table~\ref{tab:pairs} asks the question an examiner actually asks (given this
host before and after infection, which capture is worse?) and demands a margin.
The model separates the infected capture from its control by at least $\eps$ in
$28.6\%$ of pairs, never inverts by $\eps$, and is near-tied in $71.4\%$. Set
against the composition 9/4/1, this means that five of the nine strict wins,
and the single strict loss, are decided by margins below $0.05$. The model has
a direction but no usable magnitude, and a sign test on the four pairs that
clear the tolerance gives $p{=}0.0625$. \rfix{Because the checkpoint had seen
every pair, these are the most favourable numbers the representation can
produce; an out-of-sample version could only be lower.} The rule layer does no better on its
own terms: the control is ``hotter'' than its infected partner on suspect count
in $42.9\%$ of pairs and on maximum process score in $57.1\%$.

\subsection{Calibrated binary path}
\label{sec:binary}

\begin{table}[t]
\caption{Track~B on the ML corpus ($n{=}43$). Calibrated probability
separates the triage bands cleanly; agreement with manifest labels does not
follow.}
\label{tab:binary}

\setlength{\tabcolsep}{2.5pt}
\begin{tabular}{@{}lrrrr@{}}
\toprule
\textbf{Triage state} & \textbf{$n$} & \textbf{Min} & \textbf{Median} & \textbf{Max} \\
\midrule
likely benign        & 25 & 0.009 & 0.472 & 0.626 \\
high-risk ambiguous  &  1 & 0.634 & 0.634 & 0.634 \\
low-risk ambiguous   &  1 & 0.684 & 0.684 & 0.684 \\
likely malicious     & 16 & 0.757 & 0.854 & 0.978 \\
\midrule
\multicolumn{5}{@{}l}{\emph{Decisive predictions vs manifest labels ($n{=}41$)}}\\
\quad Decisive coverage $\varphi$ & \multicolumn{4}{r}{0.953} \\
\quad Precision / Recall / F1 & \multicolumn{4}{r}{0.688 / 0.440 / 0.537} \\
\quad Accuracy (selective risk $R$) & \multicolumn{4}{r}{0.537 (0.463)} \\
\quad TP / FP / TN / FN & \multicolumn{4}{r}{11 / 5 / 11 / 14} \\
\quad Pooled AUROC of $p_b$ ($n{=}43$) & \multicolumn{4}{r}{0.578} \\
\midrule
\multicolumn{5}{@{}l}{\emph{Strict-split validation ($35$ train / $8$ val)}}\\
\quad Best val.\ F1 / AUROC / PR-AUC & \multicolumn{4}{r}{0.667 / 0.438 / 0.482} \\
\quad Brier / ECE (calibrated)\rfix{$^{\dagger}$} & \multicolumn{4}{r}{0.496 / 0.505} \\
\quad Fitted temperature $T$ & \multicolumn{4}{r}{20.00 \rfix{(search bound)}} \\
\multicolumn{5}{@{}l}{\rfix{\footnotesize $^{\dagger}$on the eight validation graphs that also fitted $T$}} \\
\bottomrule
\end{tabular}
\end{table}

Track~B assigns decisive states to 41 of 43 snapshots
(Table~\ref{tab:binary}), and the calibrated probability separates the bands as
designed: \state{likely benign} rows sit at a median of $0.472$ and
\state{likely malicious} rows at $0.854$, with the two ambiguous rows inside
$[\tau_{\mathrm{low}},\tau_{\mathrm{high}}]=[0.632,0.712]$ (the high/low
qualifier refers to heuristic risk $h(x)$, not to $p_b$, which is why the
high-risk row carries the lower probability). Agreement with the manifest is
another matter. \rfix{Selective} accuracy is $0.537$ \badd{(22 of 41 decisive
snapshots; CI 0.37--0.69), or $0.512$ overall if the two abstentions are
counted as errors (22/43)}, with recall $0.440$ \badd{and specificity
$0.688$ (11/16)}, and the
dominant error cell is false negatives ($14$), that is malware-labelled
captures scored as benign.

\begin{figure*}[t]
\centering
\input{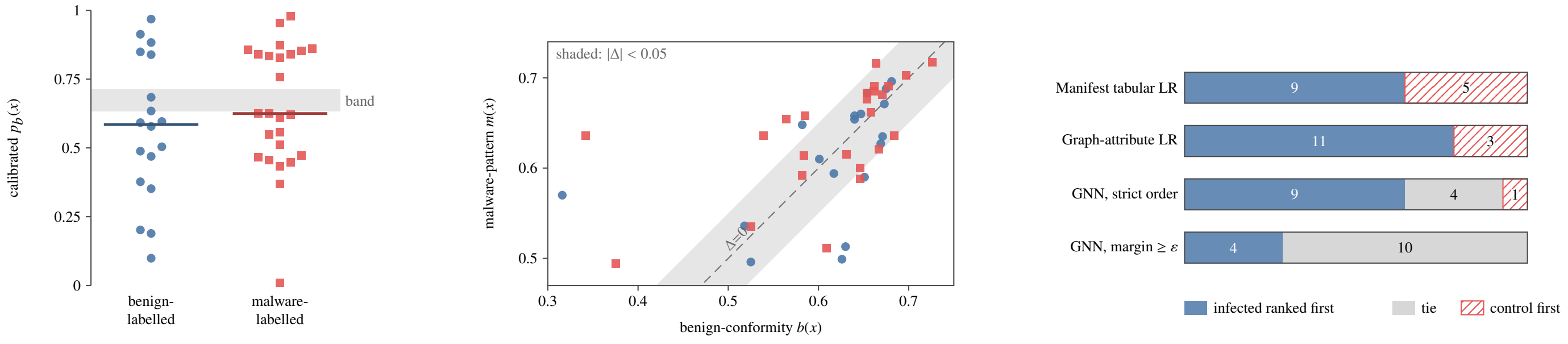}
\caption{Score fields on the ML corpus; circles are benign-labelled and
squares malware-labelled snapshots. (a)~Calibrated malware probability by
manifest label: the shaded ambiguity band is $[0.632,0.712]$ and bars are
class medians ($0.585$ vs $0.625$). (b)~Dual-model score plane on equal axes:
30 of 43 snapshots lie within $|\Delta|<0.05$ of the identity line; one
malware-labelled snapshot at $(0.000,0.011)$ falls outside the window.
(c)~Composition of the 14 matched pairs implied by
Proposition~\ref{prop:twograph} and Table~\ref{tab:pairs}: equal LOFO means
hide different behaviour, and only four GNN wins clear $\eps{=}0.05$.}
\label{fig:scores}
\end{figure*}

Figure~\ref{fig:scores}a shows why threshold tuning would not fix this. The two
label populations interleave across essentially the whole probability range:
several benign-labelled captures score above $0.9$, one malware-labelled
capture scores $0.009$, and the medians differ by $0.04$. Pooled over all 43
snapshots the calibrated probabilities give AUROC $0.578$ (exact one-sided
Mann--Whitney $p{=}0.20$), in line with the $0.615$ of the construct-validity
checkpoint. Neither can be distinguished from chance at this sample size.

Two calibration artefacts deserve comment. First, strict-split validation AUROC
is $0.438$, below chance on eight validation graphs, while best validation F1
is $0.667$. On a fold this small the two numbers are uninformative, not
contradictory, and quoting either as a headline would mislead. Second, the
fitted temperature is $20.00$, which coincides with the upper limit of the
search range. The optimum was not bracketed, so $20.00$ is a lower
bound on the temperature that minimises validation NLL, and the reported
calibration is the least softening the search allowed\rfix{; $T$ is reported
as a search bound, not as a fitted optimum}. Combined with a
calibrated ECE of $0.505$, a binned estimate computed from eight
validation graphs~\citep{silvafilho2023calibration}\rfix{, which were also
the graphs on which $T$ was fitted, so the estimate is optimistic as well as
unstable}\anote{state the number of ECE bins used for Track~B; with eight
graphs most bins are empty whatever the count}, this shows a
probability field that carries no usable confidence information. The Brier score of $0.496$ says the same thing more bluntly. A constant
prediction of $0.5$ scores $0.25$ whatever the labels, so on the validation
fold the calibrated probabilities do worse than saying nothing. We report
these figures as a diagnostic of failure and not as a calibration result\rfix{,
and we describe the probability field as diagnostically miscalibrated rather
than as calibrated or uncalibrated: eight graphs cannot establish either. A
calibration claim on this corpus would need the temperature fitted and
evaluated on disjoint folds under nested leave-one-family-out validation,
with an interval on the ECE; none of this is possible with 43 snapshots and a
single 8-graph split.}

\paragraph{Base-rate projection}
Taken at face value, the decisive verdicts have sensitivity $0.44$ (11/25; CI
0.24--0.65) and false-positive rate $0.31$ (5/16; CI 0.11--0.59), a positive
likelihood ratio of $1.41$. By Bayes' rule,
$\mathrm{PPV}=\mathrm{TPR}\,\pi/(\mathrm{TPR}\,\pi+\mathrm{FPR}(1-\pi))$~\citep{axelsson2000base}.
At the corpus prevalence of $0.61$ the precision is $0.69$; at $\pi{=}0.01$ it
is $0.014$, 70.3 false alarms per true one, and at $\pi{=}0.001$ it is
$0.0014$. A verdict is more likely right than wrong at $1\%$ prevalence
only if its likelihood ratio reaches $99$.

\subsection{Dual conformity and abstention}
\label{sec:dual}

\begin{table}[t]
\scriptsize
\caption{Track~D on the ML corpus: score ranges, routing under the
\code{calibrated} preset, gate triggers, and the gate ablation of
Section~\ref{sec:gate}.}
\label{tab:dual}

\setlength{\tabcolsep}{4pt}
\begin{tabular}{@{}lrrr@{}}
\toprule
\textbf{Score} & \textbf{Min} & \textbf{Median} & \textbf{Max} \\
\midrule
Malware-pattern $m(x)$        & 0.011 & 0.636 & 0.717 \\
Benign-conformity $b(x)$      & 0.000 & 0.646 & 0.726 \\
Margin $\Delta = m - b$       & $-0.126$ & 0.011 & 0.294 \\
\midrule
\multicolumn{4}{@{}l}{\emph{Routing (preset \code{calibrated})}}\\
\quad needs analyst review & \multicolumn{3}{r}{42 (0.977)} \\
\quad anomalous unknown     & \multicolumn{3}{r}{1} \\
\quad likely malicious      & \multicolumn{3}{r}{0} \\
\quad likely benign         & \multicolumn{3}{r}{0} \\
\midrule
\multicolumn{4}{@{}l}{\emph{Abstention triggers}}\\
\quad MC-dropout variance $+$ low margin & \multicolumn{3}{r}{28} \\
\quad MC-dropout variance only           & \multicolumn{3}{r}{12} \\
\quad Dual conflict $+$ variance $+$ margin & \multicolumn{3}{r}{1} \\
\quad Low margin only                    & \multicolumn{3}{r}{1} \\
\quad No gate fired                      & \multicolumn{3}{r}{1} \\
\midrule
\emph{Gate ablation} & \textbf{Review} & \textbf{Decisive $\varphi$} & \textbf{Gated} \\
\quad \code{disabled}    & 0.953 & 0.000 &  0 \\
\quad \code{legacy\_or}  & 1.000 & 0.000 & 43 \\
\quad \code{calibrated}  & 0.977 & 0.000 & 42 \\
\bottomrule
\end{tabular}
\end{table}

Track~D abstains on 42 of 43 snapshots (CI 0.88--1.00) and issues no decisive
verdict at all (Table~\ref{tab:dual}). Figure~\ref{fig:scores}b explains the
mechanism without reference to hyper-parameters: the malware-pattern and
benign-conformity scores are \rfix{close to} one measurement taken twice. Both occupy
overlapping bands (medians $0.636$ and $0.646$), they are \rfix{correlated
(Spearman $\rho{=}0.72$; Pearson $r{=}0.80$, which falls to $0.52$ once the
single snapshot at the origin is removed, so the rank correlation is the
figure to rely on)}, 30 of 43
snapshots lie within $|\Delta|<0.05$ of the identity line, and the margin
that the fusion layer is meant to exploit has a median of $0.011$, $2.6\%$
of its range of $0.420$. No
axis of the plane separates the labels: pooled AUROC is $0.59$ for $m$, $0.49$
for $-b$ and $0.60$ for $\Delta$.

Given Section~\ref{sec:mismatch} this outcome is coherent. The benign encoder
was trained without a single \code{clean\_benign} row, so its notion of
``normal'' was induced from captures \rfix{that the rule layer flags}.
One-class objectives presuppose uncontaminated normal data and degrade when
that assumption fails~\citep{ruff2021unifying}\rfix{; whether the controls
are in fact contaminated is unknown, but a reference set that the instrument
cannot distinguish from the positive class is unfit for a conformity model
either way}. Fitted to
ambiguous controls, the benign model learns the manifold of busy Windows
hosts, which is also where the infected captures live. \rfix{We therefore
report Track~D as a contaminated-control experiment, not as a benign
conformity model: a benign one-class path cannot legitimately be trained on
this corpus, and the architecture keeps it only so that the failure can be
measured. We do not introduce surrogate or synthetic benign rows to make the
path trainable.} In the terms of
\citet{hendrickx2024reject}, a novelty detector whose reference set already
contains the novelty has nothing left to reject on. The gates then fire as
designed (MC-dropout variance combined with low dual margin accounts for 28 of
42 abstentions), but they report a real property of the score field. They do
not manufacture caution.

\subsection{Gate ablation: abstention is not the gates' fault}
\label{sec:gate}

A natural objection is that the abstention policy is simply too conservative.
The ablation in Table~\ref{tab:dual} rules this out. With every uncertainty
gate disabled, $95.3\%$ of snapshots (41/43; CI 0.84--0.99) are still routed to
review and decisive coverage remains exactly zero, because the fused score
itself falls inside the triage band $[0.40,0.65)$ for \rfix{41 of the 43
snapshots and the remaining two are routed to \state{anomalous unknown}
before any gate (Section~\ref{sec:system}), so no snapshot receives a
decisive state under any preset}. The gates move the review rate by $2.4$ percentage points; they do not
cause the abstention. In risk--coverage terms the fusion path collapses to the
single point $\varphi{=}0$ under every preset, where selective risk is
undefined. Had we reported only the \code{calibrated} preset, the abstention
could have been put down to policy and not to representation. This is why gate
ablation belongs in the minimum reporting set for any abstaining detector.

\paragraph{Reference tabular baselines}
For orientation only (the corpus view and protocol differ, so this is
not a matched comparison), logistic regression and random forest over
manifest graph-level features on the descriptive corpus achieve F1 of $0.469$
and $0.631$ with ROC-AUC $0.489$ and $0.711$. The random forest is the
strongest single number in the MalVol-25 part of this study. We state it plainly, because
leaving out a baseline that outperforms the proposed method is exactly the
reporting practice that makes security ML results hard to
trust~\citep{arp2022dos}.

%% ==================================================================
\section[External validation on two public corpora]{External validation on two public corpora}
\label{sec:external}
%% ==================================================================

Sections~\ref{sec:mismatch}--\ref{sec:gate} rest on one dataset lineage
and 43 snapshots, and the obvious objection is that MalVol-25 is simply an
unusual corpus. We therefore repeated the audit on the two public Windows
memory corpora on which detection results are reported:
CIC-MalMem-2022~\citep{carrier2022malmem}, on which a logistic
regression has been reported at $99.97\%$ accuracy~\citep{dener2022malware}, and the memory-artefact corpus of
\citet{sihwail2021memory}. Neither releases memory images, so neither supports
the graph tracks of Section~\ref{sec:system}; both release the kind of
Volatility-derived summary that the tabular rungs of the stage ladder consume.
The question is not whether our pipeline detects malware on them. It is what
the protocol of Table~\ref{tab:protocol} reveals when it is run on corpora of
this type. Every number in this section is computed by one script from the two
public files, whose checksums are given in the data-availability statement,
with seed 2027; nothing is simulated or imputed.

\subsection[Corpora and what transfers]{Corpora and what transfers}
\label{sec:ext_setup}

\paragraph{CIC-MalMem-2022} The file holds 58{,}596 rows of 55 features
extracted with VolMemLyzer~\citep{lashkari2021volmemlyzer}, split evenly into
29{,}298 benign-labelled and 29{,}298 malware-labelled rows from fifteen
families in three categories (spyware 10{,}020; ransomware 9{,}791; trojan
9{,}487). Each malware row carries a category string of the form
\code{Category-Family-hash-$k$.raw}; each benign row carries the single word
\code{Benign}. The features are plugin summaries (process, DLL, handle,
\code{ldrmodules}, \code{malfind}, \code{psxview}, service and callback counts),
so the injection and hidden-process indicators of Table~\ref{tab:governance}
have direct counterparts; there is no network plugin, so the C2 indicator has
none.

\paragraph{The corpus of Sihwail et al.\ (2021)} The paper describes 2{,}502
malware and 966 benign samples with 8{,}898 memory features extracted from
memory images: API names resident in memory, DLLs, handles, privileges, network
connections and injected code. The file in the repository the paper names \rfix{is published as a sample
and} has
the 3{,}468 rows but 28 of the columns, all binary: nineteen API-presence
flags, seven privilege flags and the two indicators \code{Malfind} and
\code{Netscan}, which are the closest public analogues of the injection and C2
indicators used throughout this paper. We use it for the indicator-level items
of the protocol only\rfix{, and nothing in this section re-evaluates the
authors' full feature space or their memory images}.

\paragraph{What transfers} Items 1 and 2 of Table~\ref{tab:protocol} transfer
directly, with the two recorded reasons that exist in the features (a non-zero
injection count, a process absent from \code{pslist}) standing in for the rule
verdict as a \emph{surrogate} \code{label\_quality\_flag}. Item 3 does not:
neither corpus has matched controls, so there are no pairs to rank. Item 4
becomes a ladder of tabular rungs, logistic regression over all features, over
the nine rule-type indicators alone, a random forest, and the single feature
chosen on the training fold, run under three split designs of increasing
severity: stratified random folds (the design behind the published figures),
folds that hold out whole binaries, and leave-one-family-out; benign rows,
having no key, are assigned to folds at random with a fixed seed. \badd{Two
details of the grouping matter. A binary that the file lists under two
families is assigned to a single by-binary group, and when either of its
families is held out it is removed from training as well, so no test binary
has a dump in training under the by-binary or by-family designs (the leakage
row of Table~\ref{tab:malmem_ladder} is $0$ for both). The benign rows cannot
be grouped at all, so both designs hold out binaries or families on the
malware side only; what they estimate is independence from the malware
binary or family, not from the benign host, and Section~\ref{sec:ext_malmem}
reads them accordingly. All scaling and the inner temperature split are
fitted on the training fold of each design.} Item 5 loses its
point when folds hold thousands of rows, so we report fold dispersion and the
sign test over families instead. Items 6 and 7 apply unchanged to the
out-of-fold probability field: temperature fitted on an inner family-held-out
split, Brier score and ECE over fifteen equal-width bins, and the triage band $[0.40,0.65)$ of
Algorithm~\ref{alg:triage} with a two-model disagreement gate as the analogue
of binary--dual disagreement.

\subsection[CIC-MalMem-2022]{CIC-MalMem-2022}
\label{sec:ext_malmem}

\begin{table}[t]
\caption{Provenance and governance audit of CIC-MalMem-2022. The benign class
carries no provenance at all; the malware class carries a hash and a dump
index that a random split ignores.}
\label{tab:malmem_prov}

\setlength{\tabcolsep}{4pt}
\begin{tabularx}{\columnwidth}{@{}>{\raggedright\arraybackslash}Xr@{}}
\toprule
\textbf{Item} & \textbf{Value} \\
\midrule
Rows / features & 58{,}596 / 55 \\
Benign- / malware-labelled rows & 29{,}298 / 29{,}298 \\
Provenance key on benign rows & none \\
Malware families / categories & 15 / 3 \\
Distinct binaries (by hash) & 2{,}887 \\
Dumps per binary, median (range) & 10 (1--20) \\
Binaries listed under two families & 19 (366 rows) \\
Repeated (binary, dump) keys & 1{,}129 (467 byte-identical) \\
Exact duplicate feature rows & 569 (0 with conflicting class) \\
Constant / near-constant ($\le3$ values) features & 3 / 4 \\
Test malware rows whose binary is also in training, random 5-fold & 99.9\% \\
\midrule
Benign rows with $\ge1$ / $\ge3$ injected regions & 100\% / 98.3\% \\
Benign rows with a process absent from \code{pslist} & 40.7\% \\
Surrogate \code{label\_quality\_flag} on benign rows & 29{,}298/29{,}298 \\
\bottomrule
\end{tabularx}
\end{table}

\paragraph{Provenance audit} Table~\ref{tab:malmem_prov} records what the
governance schema of Section~\ref{sec:problem} can and cannot be filled with.
On the malware side the category string identifies 2{,}887 distinct \rfix{binary
identifiers (2{,}329 of SHA-256 length, 515 of MD5 length and 43 truncated to
21 or 31 characters, none a prefix of another)}, each contributing a run of dumps ($k=1,\dots,10$; median ten per binary), so
the corpus has a group structure that the usual stratified split ignores: in a random five-fold
split, $99.9\%$ of test malware rows have another dump of the same binary in
the training set. The strings also contain outright conflicts. Nineteen \rfix{identifiers,
all of SHA-256 length,} appear under two families, in every case Emotet and one of
180solutions, Pysa, CWS, Conti, Transponder or Shade, with different feature
vectors under each label; 1{,}129 (binary, dump) keys occur twice, 467 of them
as byte-identical rows. On the benign side there is nothing to audit. The
29{,}298 benign rows carry no host, session, binary or dump identifier, so they
cannot be grouped, paired or traced to a capture. The \code{benign\_subtype}
field, however, can be filled: every benign row records at least one injected
region (\code{malfind.ninjections} $\ge1$ on 29{,}298 of 29{,}298; $\ge3$ on
$98.3\%$), and $40.7\%$ record a process absent from \code{pslist}. Under the
surrogate rule the label-quality flag is set on the entire benign class (exact
95\% lower bound $0.9999$), and no row qualifies as \code{clean\_benign}. The
finding of Table~\ref{tab:governance} is therefore not a MalVol-25
peculiarity; it holds, with 1{,}953 times as many benign rows, on the
corpus the field treats as its benchmark.

\begin{table}[t]
\scriptsize
\caption{Indicator gaps on CIC-MalMem-2022 (class means; $\delta_k$ is
malware minus benign). The last column is the AUROC of the raw feature with
malware as the positive class, so values below $0.5$ mark an inverted
indicator. The nine rows above the rule are the rule-type indicators used in the
ladder of Table~\ref{tab:malmem_ladder}; those below are host-population
counts.}
\label{tab:malmem_gaps}

\setlength{\tabcolsep}{3.2pt}
\begin{tabular}{@{}lrrrc@{}}
\toprule
\textbf{Indicator} & \textbf{Benign} & \textbf{Malware} & \textbf{$\delta_k$} & \textbf{AUROC} \\
\midrule
Injected regions (\code{malfind}) & 4.96 & 9.06 & $+4.11$ & 0.279 \\
\quad medians & 4 & 3 & & \\
Unique injections & 1.37 & 2.10 & $+0.73$ & 0.152 \\
Injection commit charge & 10.3 & 1{,}928 & $+1{,}918$ & 0.195 \\
Processes absent from \code{pslist} & 1.69 & 2.07 & $+0.38$ & 0.594 \\
\quad other two \code{psxview} views & & & & 0.594 / 0.603 \\
Unlinked modules, \code{not\_in\_load} & 75.6 & 46.1 & $-29.5$ & 0.024 \\
Unlinked modules, \code{not\_in\_init} & 115.3 & 84.6 & $-30.7$ & 0.040 \\
\quad \code{not\_in\_mem} & 75.6 & 46.1 & $-29.5$ & 0.024 \\
\midrule
Kernel callbacks & 87.7 & 86.1 & $-1.7$ & 0.126 \\
Services & 394.2 & 388.5 & $-5.7$ & 0.004 \\
Processes & 42.5 & 40.3 & $-2.3$ & 0.323 \\
Handles & 12{,}119 & 8{,}398 & $-3{,}721$ & 0.017 \\
DLLs loaded & 2{,}083 & 1{,}539 & $-543$ & 0.022 \\
DLLs per process & 48.9 & 38.5 & $-10.4$ & 0.003 \\
Handles per process & 286.8 & 212.3 & $-74.4$ & 0.007 \\
\bottomrule
\end{tabular}
\end{table}

\begin{figure*}[t]
\centering
\input{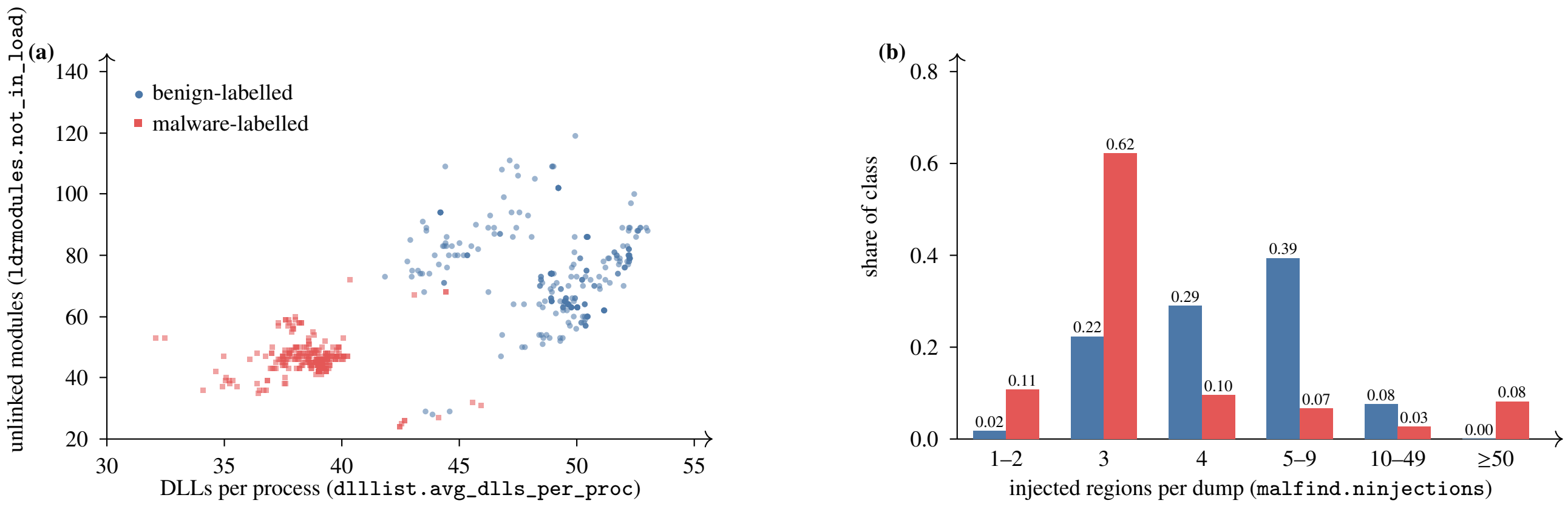}
\caption{CIC-MalMem-2022. (a)~Two host-population features for a fixed-seed
random subsample of 300 rows per class (seven of the 600 sampled rows fall
outside the plotted window and are not drawn); benign-labelled rows (circles)
are the busier hosts, and the classes barely overlap. (b)~Class-conditional share of
rows by injected-region count over all 58{,}596 rows: the benign mass sits to
the right of the malware mode at 3, and only the malware tail beyond 50 runs
the other way. A benign row shows more injected regions than a malware row in
$72\%$ of cross-class comparisons.}
\label{fig:external}
\end{figure*}

\paragraph{Indicator gaps} Table~\ref{tab:malmem_gaps} and
Figure~\ref{fig:external} repeat the second finding of
Section~\ref{sec:mismatch}. All three injection-type indicators are
rank-inverted. The raw count and the commit charge have positive mean gaps
only because of heavy malware tails (for the count, a 95th percentile of 61
and a maximum of 627 against 12 and 21 on the benign side); the median count
is 4 on benign rows and 3 on malware rows, a benign row shows more injected
regions than a malware row in $72\%$ of cross-class comparisons (AUROC
$0.279$) and more unique injections in $85\%$ (AUROC $0.152$). Of the nine
rule-type indicators only the three hidden-process views run the expected way,
and weakly ($0.59$--$0.60$); the three unlinked-module counts run strongly the
other way, with $64\%$ more \code{ldrmodules} entries absent from the load
list on benign rows. Every host-population count runs the wrong way for a
rule that reads volume as risk: the benign-labelled rows carry more services,
more processes, more handles and more loaded DLLs. Two of these counts identify the class for the great majority of rows. Seventy-three
per cent of benign rows record exactly 395 services and a further $25\%$ exactly
392, whereas $93\%$ of malware rows record exactly 389; and the mean number of
DLLs per process separates the classes with AUROC $0.9975$ (5th--95th
percentiles 43.6--52.3 on benign rows, 36.5--40.1 on malware rows), so that a
single threshold at 42.6 DLLs per process, chosen post hoc on the whole
corpus\rfix{ and reported as an exploratory illustration rather than as an
out-of-fold result}, classifies $98.7\%$ of it.
This is what a difference in host state looks like: the benign captures come
from a busier host image than the malware captures, and the label is
recoverable from how busy the host is. Which of the three causes of
Section~\ref{sec:problem} produces it cannot be decided from the features, and
without per-row provenance it cannot be decided at all.

\begin{table}[t]
\caption{Stage ladder on CIC-MalMem-2022 under three split designs (AUROC,
fold mean with the fold SD as subscript). Holding out binaries or whole families changes
nothing: the separating signal is not family-specific, and a single feature
carries all but $0.003$ of it.}
\label{tab:malmem_ladder}
\scriptsize
\setlength{\tabcolsep}{2.4pt}
\begin{tabular}{@{}lccc@{}}
\toprule
\textbf{Rung} & \textbf{Random 5-fold} & \textbf{By binary (10)} & \textbf{By family (15)} \\
\midrule
Single feature, chosen on train & $0.9975_{\pm.0003}$ & $0.9974_{\pm.0008}$ & $0.9971_{\pm.0055}$ \\
LR, 9 rule-type indicators & $0.9842_{\pm.0011}$ & $0.9842_{\pm.0036}$ & $0.9842_{\pm.0064}$ \\
LR, all 55 features & $1.0000_{\pm.0001}$ & $0.9999_{\pm.0001}$ & $0.9998_{\pm.0003}$ \\
Random forest, all 55 & $1.0000_{\pm.0000}$ & $1.0000_{\pm.0000}$ & $0.9998_{\pm.0009}$ \\
\midrule
Accuracy at $0.5$, LR / RF & 0.999 / 1.000 & 0.999 / 1.000 & 0.998 / 0.999 \\
Test binaries seen in training & 0.999 & 0 & 0 \\
Folds with AUROC $>0.5$, LR & 5/5 & 10/10 & 15/15 \\
\bottomrule
\end{tabular}
\\[2pt]
{\footnotesize The feature chosen on every training fold of every design is
\code{dlllist.avg\_dlls\_per\_proc}. Sign test over the 15 families, LR and
RF: $p=3\times10^{-5}$.}
\end{table}

\paragraph{Stage ladder under three split designs} Table~\ref{tab:malmem_ladder}
first reproduces the literature: with stratified random folds, logistic
regression reaches accuracy $0.999$ and the random forest $1.000$, in line with
the $99.97\%$ that \citet{dener2022malware} report for logistic
regression. In that design $99.9\%$ of test
binaries are also in training, so the number describes the recognition of
dumps that were seen nine times over. Removing the leakage does not lower it.
When whole binaries are held out, and when whole families are, AUROC stays at
$0.9998$ or above for both models (by-family fold SD $0.0003$ and $0.0009$),
every one of the fifteen families is ordered above chance and the sign test
gives $p=3\times10^{-5}$. The separating signal is therefore not something the
models learn from other dumps of the same binary or other members of the same
family; it is present in the benign--malware contrast itself. \rfix{These two
designs are group-independent on the malware side only. The benign rows carry
no key, so the same benign host, if there was one, sits in every training
and every test fold, and a by-family AUROC of $0.9998$ is not evidence that
the model would separate an unseen benign host from an unseen family; the
single-feature rung shows what it does separate.} Read across, the
ladder says how much of it is learned. A single feature chosen on the training
fold scores $0.997$ in all three designs, and it is the same feature every
time, the mean number of DLLs per process of Table~\ref{tab:malmem_gaps}. The
learned rungs add $0.003$ to it. The nine rule-type indicators reach $0.984$
on their own, but Table~\ref{tab:malmem_gaps} shows why: six of the nine
point the wrong way, so the regression separates the classes by weighting
them against the direction a rule layer would use.
On MalVol-25 the ladder was flat at $0.786$, the smallest value the design
could call significant; here it is flat at $1.000$. Both flat ladders mean the
same thing, that the learned component is inert, once because there is no
signal in the label and once because the signal is saturating, trivially
available and unrelated to the indicators a forensic rule would cite.

\begin{table}[t]
\caption{Calibration, abstention and base rates on the leave-one-family-out
probability field of CIC-MalMem-2022 ($n{=}58{,}596$; band $[0.40,0.65)$ of
Algorithm~\ref{alg:triage}). Every artefact that failed on MalVol-25 passes
here.}
\label{tab:malmem_cal}
\scriptsize
\setlength{\tabcolsep}{3.6pt}
\begin{tabular}{@{}lrrr@{}}
\toprule
& \textbf{LR} & \textbf{LR, $T$-scaled} & \textbf{RF} \\
\midrule
Brier score & 0.0016 & 0.0016 & 0.0011 \\
ECE (15 bins) & 0.0012 & 0.0014 & 0.0014 \\
Pooled AUROC & \multicolumn{2}{c}{0.9998} & 0.9997 \\
Median $p$, benign / malware & \multicolumn{2}{c}{0.000 / 1.000} & 0.000 / 1.000 \\
\multicolumn{4}{@{}l}{Fitted $T$ over 15 folds: 0.75--1.58, median 1.07, none at a search bound} \\
\midrule
\multicolumn{4}{@{}l}{\emph{Routing (review rate)}} \\
\quad band only, LR / RF / fused mean & \multicolumn{3}{r}{0.0003 / 0.0002 / 0.0009} \\
\quad \code{legacy\_or} (band on either, or disagree) & \multicolumn{3}{r}{0.0016 (30 + 61)} \\
\midrule
\multicolumn{4}{@{}l}{\emph{Decisive verdicts, fused mean, gates disabled}} \\
\quad Coverage $\varphi$ / selective risk $R$ & \multicolumn{3}{r}{0.9991 / 0.0011} \\
\quad TPR / FPR & \multicolumn{3}{r}{0.998 / 0.0002} \\
\quad Likelihood ratio $\mathrm{LR}^{+}$ & \multicolumn{3}{r}{5{,}844} \\
\quad PPV at $\pi=0.01$ / $0.001$ & \multicolumn{3}{r}{0.983 / 0.854} \\
\bottomrule
\end{tabular}
\end{table}

\paragraph{Calibration, abstention and base rates} Table~\ref{tab:malmem_cal}
is the mirror image of Tables~\ref{tab:binary} and~\ref{tab:dual}. On the
out-of-fold field the Brier score is $0.0016$ and ECE $0.0012$ before any
temperature is fitted; the fitted temperatures lie between $0.75$ and $1.58$,
well inside the search range, and change nothing. The medians of the two
classes sit at $0.000$ and $1.000$. The triage band catches $0.09\%$ of the
corpus, the disagreement gate adds $61$ rows, and the \code{legacy\_or} preset
still routes only $0.16\%$ to review. Decisive coverage is $0.999$, the
likelihood ratio of a decisive verdict is in the thousands, and the projected
precision at $1\%$ prevalence is $0.98$. By the argument of
Section~\ref{sec:problem}, every one of these numbers measures agreement with
provenance, and Tables~\ref{tab:malmem_prov} and~\ref{tab:malmem_gaps} say what
that provenance is: a host fingerprint. The abstention machinery is not at
fault. The gates are built to notice variance, disagreement and low margin,
and a shortcut produces none of them; as Section~\ref{sec:system} put it, an
overconfident head is not evidence of uncertainty, and the converse also holds,
a confident head is not evidence of a compromise signal. Items 6 and 7 of the
protocol, on their own, would have certified this corpus. Items 1 and 2, and
the single-feature rung of item 4, are what reveal the problem, which is why
the protocol reports them first.

\subsection[The corpus of Sihwail et al. (2021)]{The corpus of Sihwail et al.\ (2021): an indicator-level replication}
\label{sec:ext_sihwail}

\begin{table}[t]
\caption{Indicator-level audit of the public sample of \citet{sihwail2021memory}
($n{=}3{,}468$; 966 benign-labelled, 2{,}502 malware-labelled). Exact 95\%
Clopper--Pearson intervals for the benign shares are 0.50--0.57 and
0.02--0.04; $\mathrm{LR}^{+}$ and PPV treat a set flag as a decisive malware
verdict, PPV at $1\%$ prevalence.}
\label{tab:sihwail}
\scriptsize
\setlength{\tabcolsep}{3pt}
\begin{tabular}{@{}lcccc@{}}
\toprule
\textbf{Indicator set} & \textbf{Benign} & \textbf{Malware} & \textbf{$\mathrm{LR}^{+}$} & \textbf{PPV$_{0.01}$} \\
\midrule
\code{Malfind} (injection) & 0.534 (516/966) & 1.000 (2{,}502) & 1.87 & 0.019 \\
\code{Netscan} (connection) & 0.031 (30/966) & 0.552 (1{,}381) & 17.8 & 0.152 \\
Both & 0.006 & 0.552 & & \\
\midrule
\multicolumn{5}{@{}l}{Distinct feature vectors among 3{,}468 rows: 74; vectors carrying both labels: 9} \\
\multicolumn{5}{@{}l}{Family or host key: none; LR over 28 columns, stratified 5-fold: AUROC $0.917\pm0.010$,} \\
\multicolumn{5}{@{}l}{\quad accuracy $0.785$ against a majority-class $0.722$} \\
\bottomrule
\end{tabular}
\end{table}

The published sample of the second corpus supports only the
indicator-level items, and Table~\ref{tab:sihwail} gives them. The injection
indicator is set on every malware row and on $53.4\%$ of benign rows (exact
interval $0.50$--$0.57$). Taken as a detector, it has sensitivity $1.0$, a
false-positive rate of $0.53$ and a likelihood ratio of $1.87$, the same order
as the $1.41$ of Track~B on MalVol-25, and at $1\%$ prevalence its positive
verdict would be right once in fifty-four. The network-connection flag is the
informative one here: set on $3.1\%$ of benign rows and $55.2\%$ of malware
rows, it carries a likelihood ratio of $17.8$, and even so its projected
precision at $1\%$ prevalence is $0.15$. Nothing further can be audited. The
sample has no family or host key, and its 28 binary columns collapse 3{,}468
rows onto 74 distinct feature vectors, nine of which occur under both labels,
so a stage ladder or a permutation null would measure the two indicators and
little else. The one number we report from a model, a logistic regression at
AUROC $0.917$ and accuracy $0.785$ against a majority-class $0.722$, says as
much.

\subsection[Cross-corpus synthesis]{Cross-corpus synthesis}
\label{sec:ext_synthesis}

\begin{table*}[t]
\caption{The seven reporting artefacts of Table~\ref{tab:protocol} on three
corpora. The learned-metric items (4--7) range from failing to passing
perfectly; the label-quality items (1--2) fail on all three.}
\label{tab:cross}
\scriptsize
\setlength{\tabcolsep}{4pt}
\begin{tabularx}{\textwidth}{@{}>{\raggedright\arraybackslash}p{0.17\textwidth}>{\raggedright\arraybackslash}X>{\raggedright\arraybackslash}X>{\raggedright\arraybackslash}X@{}}
\toprule
\textbf{Artefact} & \textbf{MalVol-25, ML view ($n{=}43$)} & \textbf{CIC-MalMem-2022 ($n{=}58{,}596$)} & \textbf{Sihwail et al.\ 2021, sample ($n{=}3{,}468$)} \\
\midrule
1. Label-quality fields & 18/18 benign rows flagged, 0 train-eligible; paired controls with host provenance & surrogate flag on 29{,}298/29{,}298 benign rows; no benign provenance key; 19 binaries under two families; 467 duplicate rows & \code{Malfind} set on 516/966 benign rows; no key; 74 distinct vectors, 9 under both labels \\[1pt]
2. Mismatch rate and indicator gaps & $\mu\ge0.73$; injections $-0.80$, C2 $-1.93$ (inverted) & injections rank-inverted (benign $\succ$ malware in 72\% of comparisons, 85\% for unique injections); services, DLLs, handles, unlinked modules all higher on benign rows & injection flag: $+0.47$ in share; connection flag: $+0.52$ \\[1pt]
3. Within-pair rank diagnostics & rank accuracy 0.286, near-tie 0.714, inversion 0 \rfix{(frozen checkpoint, in-sample)} & no matched controls: ten dumps per binary, none clean & not applicable \\[1pt]
4. Stage ladder & $0.643\to0.786\to0.786$; $\Delta$(GNN $-$ attribute LR) $=0.000$ \rfix{(GNN rows in-sample)} & $0.997$ (one feature) $\to0.984$ (indicators) $\to1.000$ (LR, RF); flat across random, by-binary and by-family splits & LR over 28 columns 0.917; two indicators carry it \\[1pt]
5. Null, fold count, design power & $p=0.044$ at $B=500$ \rfix{(across pairs, rescored checkpoint)}; $0.786$ is the smallest significant mean \rfix{of a decided design; $p=0.011$ with ties dropped, $0.212$ with ties as losses} & 15/15 families above chance, $p=3\times10^{-5}$; fold SD $\le0.001$ & no groups, no null \\[1pt]
6. Temperature, Brier, ECE & $T=20.00$ at the bound; $0.496$; $0.505$ & $T\in[0.75,1.58]$; $0.0016$; $0.0012$ & not supported \\[1pt]
7. Review rate, coverage, gate ablation, PPV at 1\% & $0.977$; $0$; $0.953$ with gates off; $0.014$ & $0.0009$; $0.999$; $0.0016$ with gates on; $0.983$ & flag as verdict: PPV $0.019$ (injection), $0.152$ (connection) \\
\midrule
Reading & a silent model on an unverified benign class & a confident model on a host fingerprint & instrument noise on half of the benign class \\
\bottomrule
\end{tabularx}
\end{table*}

Table~\ref{tab:cross} places the three corpora side by side. In none of
them does the benign class behave like a verified benign class: on MalVol-25
every control is flagged and two headline indicators are inverted; on
CIC-MalMem-2022 every benign row carries injected regions, two injection
indicators are rank-inverted, and the benign rows are the busier hosts; on the
Sihwail sample the injection indicator fires on half the controls. What differs
is what the learned-metric artefacts say. On MalVol-25 they fail, and the model
is silent. On CIC-MalMem-2022 they pass without exception, and the model is
confident, because one host-population feature separates the classes on its
own and survives every hold-out we can construct. The two outcomes are
opposite ends of the same defect. A corpus whose benign class is asserted by
provenance can carry no label-related signal at all, or it can carry a signal
that is entirely about provenance; in both cases a detector trained on it is
right about the data and wrong about the world, and in neither case do
calibration, coverage or a base-rate projection notice. This is the argument
for the order of the protocol. Items 1 and 2 are cheap, they need no model, and
on all three corpora they are the items that fail.

%% ==================================================================
\section{Discussion: a validation protocol}
\label{sec:discussion}
%% ==================================================================

\begin{table*}[t]
\caption{Reporting protocol for learned forensic triage on corpora whose
ground truth is asserted rather than verified. Each item is a falsifiable
artefact, not a narrative caveat.}
\label{tab:protocol}

\setlength{\tabcolsep}{3.5pt}
\begin{tabularx}{\textwidth}{@{}>{\raggedright\arraybackslash}p{0.635\columnwidth}>{\raggedright\arraybackslash}X@{}}
\toprule
\textbf{Report} & \textbf{Rationale} \\
\midrule
Per-row label-quality fields & Distinguishes a verified benign class from an
asserted one; makes strict-training eligibility auditable. \\[1pt]
Mismatch rate $\mu$ and indicator gaps $\delta_k$, with the AUROC of
each raw indicator & Reveals mismatch
directly: here $\mu\ge0.73$ and two of three gaps are inverted; on
CIC-MalMem-2022 the injection indicators are rank-inverted while one
host-population count reaches AUROC \rfix{$0.997$ under leave-one-family-out
validation}. \\[1pt]
Within-pair rank accuracy, near-tie and inversion rates at a stated $\eps$ &
Aggregate AUROC over matched pairs is dominated by between-family variance
that does not transfer. \\[1pt]
Stage ladder against a feature-only baseline, with W/T/L\rfix{, each rung
marked as refitted per fold or scored by a frozen checkpoint}, under
random, by-binary and by-family splits & Separates
learned structure from re-expression of the rules that curated the corpus;
equal means can hide different compositions; a ladder that is flat at
$1.0$ is as diagnostic as one that is flat at $0.786$\rfix{; an in-sample
diagnostic must not be read as an out-of-sample estimate}. \\[1pt]
Permutation null with $B$, fold count, SD, Monte-Carlo error\rfix{, tie
handling} and
design power & A mean over 14 two-graph folds is not an effect size;
$0.786$ is the smallest \rfix{mean a decided design of 14 pairs} can call
significant\rfix{, and a significance that disappears when ties count as
losses is not one}. \\[1pt]
Fitted temperature, Brier, ECE & A boundary temperature with ECE near $0.5$
signals an uninformative probability field. \\[1pt]
Review rate, decisive coverage, gate ablation, base-rate PPV &
Distinguishes conservative policy from a degenerate score field, and prices
the examiner's workload. \\
\bottomrule
\end{tabularx}
\end{table*}

Table~\ref{tab:protocol} collects the instrumentation into a protocol. All
seven items were produced by scripts that were already needed to run the
pipeline, and each can falsify a claim the paper might otherwise have made.

\badd{\paragraph{What the evidence establishes, and what it does not} The
evidence establishes four things. On MalVol-25 a transparent rule instrument
contradicts the benign label on every control and two headline indicators run
the wrong way; on CIC-MalMem-2022 every benign row carries injected regions
and the classes are separated by host-population counts that no forensic rule
would cite; on the Sihwail sample the injection indicator fires on half the
benign class; and on MalVol-25 a graph model that has seen every fold cannot
order the pairs by a usable margin and the fusion path never decides. The
evidence does not establish that any control was compromised: $\mu$ measures
discordance between two imperfect witnesses, the three causes of
Section~\ref{sec:problem} are not apportioned, and the rule layer's own error
rates against host state are unknown. Nor does it establish that the graph
network cannot learn; it establishes that this corpus cannot show that it has.
The CIC-MalMem-2022 result is the sharper lesson for the field: a model can be
calibrated, decisive and correct on every held-out family and still be
scientifically invalid as a detector, because what it has learned is the
capture condition. A high score is not a reason to trust a model; a verified
benign class is.}

\paragraph{Silence and shortcut}
Section~\ref{sec:external} shows that the protocol has two failure modes to
catch, not one. On MalVol-25 the learned component is \emph{silent}: no rung
of the ladder is separable from the resolution of the design, the probability
field carries no confidence information and the fusion path abstains. On
CIC-MalMem-2022 it is \emph{confident}: the ladder is flat at $1.0$ under
every split, the field is calibrated and decisive, and abstention is
negligible. The label-quality items fail on both, and for the same reason,
that the benign class was asserted rather than verified. A reviewer who saw
only items 4 to 7 would reject the first corpus and accept the second; items
1 and 2 show that the second is the more dangerous of the two, because its
detector will be deployed. The single-feature rung is the cheapest guard
against the shortcut: if one plugin count matches the full model, the model
has learned the capture condition, whatever else it has learned.

\paragraph{Label provenance is a system artefact, not a caveat}
Once \code{train\_eligible=False} on 18 of 18 benign rows is machine-readable,
a strict trainer cannot silently proceed, and a reader can verify in one query
that no clean benign manifold existed. For corpus builders the implication is
concrete. A benign baseline should be verified, not merely captured
before detonation: built from known-good media, with loaded images allow-listed
against reference hashes, resident security tooling enumerated, and acquisition
tool and timing recorded. The outcome should ship as per-row fields in the
dataset's documentation~\citep{grajeda2017datasets}. CIC-MalMem-2022
shows what their absence costs: its benign rows carry no provenance at all, so
the corpus cannot be grouped, paired or audited on the benign side, and the
host-state difference that separates its classes can neither be explained nor
removed.
This is in line with recent work that makes the origin of the data part
of a forensic dataset's description~\citep{breitinger2023sharing} and ties data
set creation to stated principles and use cases~\citep{gobel2025datasets}.
Verifying the clean half is the item we would add to those lists.

\paragraph{Telling the three causes apart}
The three causes of Section~\ref{sec:problem} call for different remedies, and
ordinary forensic means are enough to separate them. \rfix{Acquisition
artefacts are excluded when the guest is paused and its memory is read by the
hypervisor: such a snapshot is atomic in the sense of
\citet{pagani2019temporal}, every page reflecting one instant, and
\citet{ottmann2024inconsistencies} report snapshots of frozen systems as
consistent in their measurements. Atomicity is a property of the copy, not of
the state copied: a kernel structure that was mid-update at the instant of
the pause is captured mid-update, so a paused snapshot removes the acquisition
cause and leaves the other two untouched. Live acquisition with an in-guest
tool, by contrast, adds inconsistencies of its own~\citep{rzepka2024causal},
and the dataset descriptor does not say which method produced the dumps.}
Instrument noise can be
measured by attributing every flagged region, hook and connection on a baseline
to the module that owns it and checking that module against known-good hashes.
What is left after allow-listing is the residue that needs an explanation.
\badd{Allow-listing has a known blind spot: it clears what it recognises, and
a region attributed to a signed module can still have been modified in
memory~\citep{block2023imgmalfind}, so the residue is a lower bound on what
needs explaining. Repeated baseline captures of the same guest add a second
check: an indicator that varies between two captures of an unchanged host is
instrument or acquisition noise, one that is stable is a property of the
host.}
Latent contamination is tested by rebuilding the guest offline from known-good
media and capturing again. A benign memory baseline should therefore carry, per
row, its build source and image hash, the acquisition method (tool, version,
guest paused or live), uptime at capture, the resident security tooling, the
share of loaded images and executable private regions attributed to known-good
hashes, and the rule verdict after allow-listing with the name of the reviewer.
Each field is cheap to produce while a corpus is being built, and none can be
reconstructed afterwards.

\paragraph{Paired controls demand rank diagnostics}
Aggregate AUROC over matched pairs is dominated by between-family variance,
which is exactly the variance that does not transfer. The within-pair questions
are harsher and closer to practice, and low rank accuracy together with zero
inversions is diagnostic in itself: the model is not wrong, it is
silent, and no aggregate metric expresses that.

\paragraph{Abstention rate is a result, not a safety feature}
A system that routes $97.7\%$ of inputs to a human has not achieved safety. It
has moved the workload. Because abstention is easily presented as conservatism,
review rate, decisive coverage and a gate ablation should be reported together;
here that triple is what separates ``cautious policy'' from ``degenerate score
field''.

\paragraph{Hybrid rule--learning systems leak by construction}
When the same rule engine both curates the corpus and supplies graph-level
features, apparent learning may be re-expression~\citep{kapoor2023leakage}.
Equal LOFO means
for attribute-only regression and the GNN, and a $0.000$ delta between pooling
variants, are \rfix{consistent with} re-expression\rfix{; they do not prove it,
because the profile ablation that would separate the two readings was not
run}. A learned model that
adds $0.003$ to a single plugin count, as on CIC-MalMem-2022, is the same
finding from the other side. \rfix{The leakage-controlled comparison that the
architecture was built to support, and that this study did not complete, is a
ladder with nine rungs under one protocol: the rule layer alone; logistic
regression and a random forest over manifest features; a single-feature
shortcut baseline chosen on the training fold; the GINE model under the
\code{full}, \code{no\_manifest\_leakage} and \code{structure\_only} profiles,
each refitted without the held-out family; suspect-mask pooling; and the
within-pair diagnostics for every rung. Preprocessing, feature selection,
early stopping, the temperature and every threshold must be fitted inside the
training fold. Until that ladder is run on a corpus with verified benign rows,
the leakage finding of this paper is the weaker one: equal means and a zero
pooling delta, not a demonstrated absence of structural information.}

\paragraph{Evidential weight}
Evaluative opinions weigh how probable the observations are under competing
propositions~\citep{casey2020evaluative}, and the output of an automated system
can enter that reasoning only once its performance has been
characterised~\citep{bolle2020evaluations}. For Track~B the characterisation is
a likelihood ratio of $1.41$: a \state{likely malicious} verdict makes
compromise $1.41$ times more probable than before the tool ran, which is
negligible support.

\paragraph{Implications for practice}
For an examiner, the rule layer and its evidence pointers remain usable
whatever the learned components do, which is why the architecture keeps them
independent. The learned scores should not yet inform a triage decision: a
near-tie rate of $0.71$ is an error characteristic that a validation report
would have to disclose~\citep{horsman2019tool}\rfix{, and a system that abstains
on 42 of 43 inputs has not demonstrated operational triage at all; it has
demonstrated that its score field cannot be thresholded}. Nothing here argues
against learning as such. Where ground truth is adjudicated at artefact level
and the decision process is explicit, as in the Registry triage of
\citet{ghanem2026winregrl}, the same questions can be answered with more
confidence than snapshot labels allow.

\badd{\paragraph{A paired corpus that would support a detection claim} The
experiments this paper could not run define the corpus that would allow them,
and we set it out as a proposal, not as work done. Each family contributes a
clean baseline and a matched infected capture of the same guest, with the
baseline verified as described above: built from known-good media with the
image hash recorded, loaded modules and executable private regions attributed
to allow-listed hashes, resident security tooling enumerated, the rule layer
run after allow-listing and its residue adjudicated by a named examiner.
Acquisition is by hypervisor snapshot of the paused guest, with the tool,
version and uptime recorded, and each baseline is captured at least twice so
that instrument and acquisition noise can be separated from host state. The
design should span at least two Windows releases, two host configurations
(with and without a third-party security product) and two acquisition methods
(paused snapshot and in-guest live tool) on a subset, so that each factor's
contribution to $\mu$ can be estimated. The sample size follows from
Section~\ref{sec:construct}: 28 pairs give 80\% power against a model that
orders three pairs in four correctly, and a corpus meant to detect a weaker
effect needs more. The analysis plan is fixed in advance: per-rung refitting
without the held-out family, exact enumeration of the within-pair null, ties
reported under both conventions, nested calibration, the nine-rung ladder
above, and every artefact of Table~\ref{tab:protocol} published with the
corpus. Essential to a detection claim are the verified baselines, the
per-fold refit and the leakage profiles; the repeated captures and the
acquisition-method arm are what would let the three causes be apportioned.}

%% ==================================================================
\section{Limitations and threats to validity}
%% ==================================================================

\emph{External validity} is the binding constraint: $n{=}30$ and $n{=}43$
snapshots over 15 families, one dataset lineage, virtualised guests. Nothing
here characterises graph learning on memory forensics in general. It
characterises what can be measured on a small paired corpus without verified
benign ground truth. \rfix{The GNN results in particular are in-sample
diagnostics of one frozen checkpoint, so no statement about generalisation to
an unseen family is made or available; the per-fold refit that would provide
one was not run, and neither were the two leakage-controlled profiles.} The external corpora of Section~\ref{sec:external}
widen the claim about corpora, not the claim about the pipeline: they release
plugin summaries, not images, so the graph tracks were not run on them, the
label-quality flag there is a surrogate built from two recorded reasons
rather than the rule layer's verdict, and the split designs could group only
the malware side because the benign rows carry no key. The Sihwail sample is a
column subset of its corpus and supports indicator-level claims only. We do not
know how the two classes of CIC-MalMem-2022 were captured, so the host-state
difference we measure is a finding about the file, not an explanation of it. \emph{\rfix{Statistical conclusion} validity} is correspondingly weak: every
interval is wide, LOFO folds contain two graphs, the design detects a
model that orders three pairs in four correctly with probability $0.52$, and
Table~\ref{tab:ladder} is ordinal, not an effect size\rfix{; the one
significant GNN value depends on dropping four saturated ties, the permutation
null was drawn across pairs and rescored a fixed checkpoint, and the Track~B
calibration figures come from eight graphs that also fitted the temperature}. \emph{Construct
validity} is the subject of the paper. Folder provenance is an imperfect
operationalisation of compromise, and the rule layer is itself an unvalidated
instrument, so $\mu$ measures disagreement between two imperfect witnesses and
not contamination. We cannot say which of the three causes of
Section~\ref{sec:problem} dominates; doing so needs baselines whose cleanliness
has been established independently\rfix{, and the title of the paper is
chosen to say exactly that: the benign class is unverified, which is all the
evidence shows}. \emph{Internal validity} risks include
schema drift between manifest versions and artefacts of graph construction,
which health flags and artefact contracts mitigate but do not remove\rfix{;
the health flags certify parseable output, not correct output}. \badd{\emph{Acquisition} is a
limitation in its own right: the descriptor records neither the acquisition
tool nor whether guests were paused, so the share of indicators that the
capture itself produced is unknown for every row, and the chain of custody
of the public images is the dataset's, not ours.} Derived
statistics rest on values reported to three decimals, adversarial robustness is
not evaluated, and all results are inference-time evaluations of frozen
checkpoints. \badd{\emph{Independent ground truth} is absent by construction
on all three corpora, and no analysis in this paper can supply it; the
proposal of Section~\ref{sec:discussion} is the only route to it.}

%% ==================================================================
\section{Conclusion}
%% ==================================================================

We set out to build a graph-based malware triage system over Windows memory
forensics and ended up with a validation study, because the corpus would not
support the claim the architecture was designed to make. Every benign-labelled
snapshot carries a label-quality flag\rfix{; that is, a transparent instrument
contradicts the label on every control, which shows the benign class to be
unverified and no more}. Injection and C2 means run higher on the
benign side. The learned model matches rule-attribute logistic regression in
mean, differs from it in fold composition and is invariant to pooling\rfix{,
and even with every pair in its training set} it
separates an infected capture from its own control by a meaningful margin in
$28.6\%$ of pairs; and the fusion layer abstains on $97.7\%$ of inputs for
reasons that survive the removal of every uncertainty gate. Each statement is a
measurement, produced by instrumentation that a detection-focused pipeline
would have had no reason to build. The audit does not depend on the
corpus. On CIC-MalMem-2022 the benign class is flagged in its entirety, the
injection indicators are rank-inverted, and the $99.9\%$ accuracy of the
literature reduces to one host-population feature that survives every
hold-out \rfix{the file allows}; on the sample of Sihwail et al.\ (2021) the injection indicator
fires on half of the benign class. A silent model and a confident model are
two outcomes of the same unverified benign class, and only the label-quality
items of the protocol tell them apart from a working detector. \rfix{The
methodological contribution is therefore not a detector but a way of
reporting: label-quality fields, the mismatch rate and indicator gaps, fold
compositions with their tie convention, the in-sample or out-of-sample status
of every rung, and the abstention rate with its ablation, as artefacts a
reviewer can check.} Three pieces of work follow directly:
a paired memory corpus with independently verified baselines and at least the
28 pairs that Section~\ref{sec:construct} asks for, \rfix{a per-fold refit of
the graph model with} exact enumeration of the
within-pair null in place of Monte-Carlo permutation, and \rfix{the first run
of} the leakage-controlled profiles on that corpus. Learned forensic triage needs
corpora whose benign class has been verified, not asserted. Until it has them,
the responsible contribution is to measure the gap and report it.

%% ==================================================================
\section*{Data availability}
The pipeline, manifests, analysis files, and the derived-statistics scripts will
be released on acceptance\badd{, as one archive containing: the extraction and
graph-construction code with the Volatility~3 release and symbol-table source
pinned; the graph schema and the rule-layer configuration (indicator weights
and band thresholds); the model configurations, random seeds and the frozen
checkpoints behind every reported number; the two manifest views with the
governance fields; the construct-validity report and the per-sample score
files from which Figure~\ref{fig:scores} is drawn; the derived-statistics,
external-audit and figure scripts with an environment file; SHA-256 checksums
of every released file; and a README that reproduces each table from the
archive}\anote{confirm that each listed item exists and will be released; remove
any that will not}. The
underlying memory images are public~\citep{dunsin2025malvoldata} at \url{https://doi.org/10.21227/kg5b-nf37}.
The external audit of Section~\ref{sec:external} uses two public files as
distributed by their authors, \code{Obfuscated-MalMem2022.csv} (58{,}596 rows,
MD5 \code{9e1be3f2b0393d5d287dae5d529f356a}) and
\code{memory\_dataset\_sample.csv} from the repository named in
\citet{sihwail2021memory} (3{,}468 rows, MD5
\code{2ab6bf8a7355cef62d83912bb74d7ce7}); the script that computes every
number in that section and regenerates Figure~\ref{fig:external} accompanies
the artefacts.

\section*{CRediT authorship contribution statement}
\textbf{Viraj Rajeshbhai Chudasama:} Conceptualization, Methodology,
Software, Investigation, Formal analysis, Writing -- original draft.
\textbf{Eduardo Palmieri:} Resources, Data curation, Validation, Writing --
review \& editing. \textbf{Dipo Dunsin:} Resources, Data curation,
Validation, Writing -- review \& editing. \textbf{Aisha Junejo:} Validation,
Writing -- review \& editing. \textbf{Mohamed Chahine Ghanem:}
Conceptualization, Methodology, Supervision, Project administration, Writing
-- review \& editing.

\section*{Declaration of competing interest}
The authors declare that they have no known competing financial
interests or personal relationships that could have appeared to influence the
work reported in this paper. Three of the authors are also authors of the
MalVol-25 dataset on which part of the study is conducted; the dataset is
public and is cited in the third person.

\section*{Funding}
This research did not receive any specific grant from funding agencies
in the public, commercial, or not-for-profit sectors.

\section*{Declaration of generative AI and AI-assisted technologies in the
manuscript preparation process}
During the preparation of this work the authors used Claude
(Anthropic) for language editing\rfix{,} \LaTeX{} formatting \rfix{and the
drafting of the derived-statistics and external-audit scripts described in
Section~\ref{sec:setup}, which the authors reviewed, ran and checked}. After using this tool, the authors reviewed and edited the content as needed and take full responsibility for the content of the publication. No experimental result was invented, extrapolated or altered by a generative tool.

\scriptsize
\bibliographystyle{elsarticle-harv}
\bibliography{refs}

\end{document}